\documentclass[aps,prl,nofootinbib,twocolumn,superscriptaddress]{revtex4-1}
\usepackage{euscript}
\usepackage{amssymb}
\usepackage{amsfonts}
\usepackage{amsbsy}
\usepackage{epsfig}
\usepackage{subfigure}
\usepackage{amsthm}
\usepackage{amscd}
\usepackage{amstext}
\usepackage{verbatim}
\usepackage{amsmath}
\usepackage{url}
\usepackage{bm}
\usepackage[pdftex]{hyperref}
\usepackage{color}

\usepackage{graphicx}

\usepackage{dcolumn}
\begin{document}
\title{Dynamical behaviour of soliton train in holographic superfluids at zero temperature}
\author{Meng Gao}
\email{gm@mail.bnu.edu.cn}

\affiliation{School of Physics and Astronomy, Beijing Normal University, Beijing 100875,
	China}
\affiliation{Key Laboratory of Multiscale Spin Physics, Ministry of Education, Beijing Normal University, Beijing 100875, China}
\author{Xin Li}
\email{xin.z.li@helsinki.fi}

\affiliation{Department of Physics, University of Helsinki,P.O. Box 64, FI-00014 Helsinki, Finland}
\affiliation{Helsinki Institute of Physics, University of Helsinki, P.O. Box 64, FIN-00014 Helsinki, Finland}
\author{Yu Tian}
\email{ytian@ucas.ac.cn}

\affiliation{School of Physical Sciences, University of Chinese Academy of Sciences,
	Beijing 100049, China}
\affiliation{Institute of Theoretical Physics, Chinese Academy of Sciences, Beijing
	100190, China}
\author{Peng Yang}
\email{pengyang23@sjtu.edu.cn}

\affiliation{School of Physics and Astronomy, Shanghai Jiao Tong University, Shanghai 200240, China}
\affiliation{Wilczek Quantum Center, School of Physics and Astronomy, Shanghai Jiao Tong University, Shanghai 200240, China}
\affiliation{Shanghai Research Center for Quantum Sciences, Shanghai 201315, China}
\author{Hongbao Zhang}
\email{hongbaozhang@bnu.edu.cn}

\affiliation{School of Physics and Astronomy, Beijing Normal University, Beijing 100875,
	China}
\affiliation{Key Laboratory of Multiscale Spin Physics, Ministry of Education, Beijing Normal University, Beijing 100875, China}

\smash{\hspace{1.1\textwidth}%
	\raisebox{-\height}{\includegraphics[height=1cm]{example-image-a}}%
	\hspace{-0.9\textwidth}%
	\textbf{HIP-2024-25/TH}%
}

\begin{abstract}
We construct a soliton train configuration with holographic superfluid
model under AdS soliton background. We investigate the stability of a soliton train using Bloch waves under two distinct quantization schemes. Upon imposing a minor perturbation on the soliton train system, it has been observed that there exist two elastic modes and one phonon mode. Most importantly, we find, under soliton train background, that there is a rich phase diagram concerning the chemical potential under standard quantization. Nevertheless, there are no unstable modes under alternative quantization.
\end{abstract}
\maketitle

\section{Introduction and motivation}

Recently, researchers have shown a growing interest in multiparticle systems due to their fascinating emergent dynamical behaviors. Among these systems, superfluidity has consistently been a topic of great interest.
Various collective excited modes can be observed in strongly coupled quantum systems. For example, soliton and quantum vortex have been studied in superfluids both experimentally \cite{lixin-1,lixin-2} and theoretically. In particular, matter-wave soliton can be engineered in atomic superfluids using cold-atom technology, allowing direct observation of their movement \cite{collective-13,key-4,key-5,key-6,key-7}.
Nevertheless, it is widely recognized that the physical mechanism cannot be adequately elucidated using traditional perturbation methods because of its strong coupling nature in superfluids.
Fortunately, holographic duality offers an explanation of strongly coupled quantum systems by means of weakly coupled gravitational theories in a higher-dimensional space \cite{Susskind,Maldacena,Witten}. In this context, the AdS/CFT duality serves as a fundamental illustration of the holographic principle. It posits that the gravitational system within anti-de Sitter space is dual to a conformal field theory existing on its conformal boundary. Some works have already been performed on the vortex configuration and its dynamics in superfluid via AdS/CFT duality \cite{key-9,key-10,key-11,key-12}. Previous theoretical studies have illuminated the behavior of the individual soliton in holographic superfluids \cite{Guo,gaomeng}, however, the static configuration and stability of the soliton train have not been studied via AdS/CFT duality. We anticipate some fascinating behavior concerning the soliton train dynamics. Understanding the behavior of these collective modes is essential for investigating non-equilibrium dynamics \cite{key-15}.

A soliton is a localized wave packet that can propagate with constant intensity because of a balance between dispersion and nonlinear interactions. Soliton trains are formed by a series of interacting solitons arranged in sequence. Soliton trains have been observed experimentally \cite{Joshua,KEStrecker} and can be constructed by modulational instability \cite{Nguyen}. In soliton train configurations, solitons are well separated from each other due to repulsive interactions \cite{Nguyen}. In this paper, we are interested in a holographic soliton train at zero temperature, including both its static configuration and stability.
The Abelian-Higgs model in the AdS soliton background enables the realization of a holographic setup at zero temperature \cite{Nishioka}. The motivation of this letter is to explore the behavior of soliton trains in two different types of superfluid systems. In fermionic cold-atom systems, there exist two well-known types of superfluids, that is, BEC superfluids and BCS superfluids. There exists an intermediate region between the two types of superfluids, which is known as the BEC-BCS crossover \cite{M_Randeria,Timmermans}.
Soliton train configurations have been studied using the Bogoliubov-de Gennes (BdG) equations \cite{key-16}. However, in this study, we aim to create soliton trains within two different types of superfluid system and analyze their dynamical behavior separately. The two distinct superfluid systems can be achieved through two types of quantization schemes, that is, standard quantization and alternative quantization, corresponding to BCS-like superfluid and BEC-like superfluid, respectively \cite{gaomeng}. Additionally, by observing their distinct manifestations, we can differentiate between the two types of superfluids and proceed to investigate the BEC-BCS crossover. The chemical potential $\mu$ and perturbation mode $k$ play crucial roles in influencing the dynamical phase transition of a soliton train. For the parameter of chemical potential, the background configuration becomes unstable and evolves to a homogeneous superfluid phase beyond a certain critical value. For the wave vector, there exists only one unstable mode under perturbation for the vanishing value of wave vector $k$; however, the other unstable mode arises as $k\neq 0$. As such, the chemical potential and wave vector will govern the life trajectory of the soliton train.
In this context, we uncover the wealth of pattern of dynamical phase transition for standard quantization. Furthermore, we identify three types of modes that produce varying effects on the soliton trains. The contents mentioned above can be realized through experiments.

The paper is structured as follows. In the next section, we introduce the zero-temperature holographic superfluid model. In Section III, we initially present the static solutions of soliton train under two quantization schemes. In Section IV, we will conduct a linear analysis about soliton train system, named quasi-normal mode (QNM), to study the dynamic behavior of soliton train. Finally, in Section V, we draw our conclusions and provide some outlook for future research.

\section{Holographic setup}

The simplest holographic model to depict the two-dimensional superfluid relies on the Abelian-Higgs model \cite{S.A,C.P}. This model involves a complex
scalar field $\Psi$ coupled to a $\left(U\left(1\right)\right)$ gauge field $A_{a}$
within the gravitational framework of dimensions $\left(3+1\right)$, with a negative cosmological constant
related to the AdS radius $L_{AdS}$ as $\Lambda=-3/L_{AdS}^{2}$, and the associated action is given by
\begin{equation}
S=\int_{\mathcal{M}}d^{4}x\sqrt{-g}\left[R+\frac{6}{L_{AdS}^{2}}+\frac{1}{e^{2}}\mathcal{L}_{matter}\right].\label{eq:HHH model}
\end{equation}
Where, the Lagrangian for matter fields reads
\begin{equation}
\mathcal{L}_{matter}=-\frac{1}{4}F_{ab}F^{ab}-\left|D\Psi\right|^{2}-m^{2}\left|\Psi\right|^{2},\label{eq:Lagrangian}
\end{equation}
where $D_{a}=\nabla_{a}-iA_{a}$, $\nabla_{a}$ represents the covariant
derivative that is compatible with the metric. Here, $e$ and $m$ are the charge
and mass of the complex scalar field $\Psi$, respectively. For simplicity, we will work in the probe limit, namely the backreaction
of the matter fields to the spacetime is neglected, which can be accomplished by taking the
limit $e\rightarrow\infty$.
 We are intrigued by the superfluid system at zero temperature; therefore, we will consider the AdS soliton metric \cite{Guo,Guo2,Nishioka},
\begin{equation}
ds^{2}=\frac{1}{z^{2}}\left[-dt^{2}+\frac{dz^{2}}{f(z)}+dx^{2}+f(z)d\xi^{2}\right]\label{eq:metric}
\end{equation}
as our background geometry, where the AdS radius $L_{AdS}$ has been set to one and $f(z)=1-(\frac{z}{z_{0}})^{3}$. The spacetime geometry ends at $z_{0}$. To ensure a smooth geometry at the tip, we need to impose a periodicity of $\frac{4\pi z_{0}}{3}$ to the $\xi$ coordinate. For ease of numerical calculations, we shall assume $z_{0}=1$ below.

Variation of the action gives rise to the equations of motion for the matter fields, which can be written as
\begin{equation}
\nabla_{a}F^{ab} =J^{b},\quad
D_{a}D^{a}\Psi-m^{2}\Psi =0\label{eq:Klein-Gordon equation}
\end{equation}
with $J^{b}=i[\Psi^{*}D^{b}\Psi-\Psi\left(D^{b}\Psi\right)^{*}]$.
The asymptotic behavior for the bulk fields near the AdS boundary
goes as \cite{C.P,Kovtun}
\begin{equation}
A_{\mu}=a_{\mu}+b_{\mu}z+\cdots,\quad
\Psi=\Psi_{-}z^{\Delta_{-}}+\Psi_{+}z^{\Delta_{+}}+\cdots.\label{eq:psi asymptotic solution}
\end{equation}
The mass of scalar field is related to the conformal dimension $\Delta$
of the condensate as $\Delta_{\pm}=3/2\pm\sqrt{9/4+m^{2}}$ \cite{Klebanov}, whereby the mass is required to satisfy
$m^{2}\geq-9/4$, which is referred to as the Breitenlohner-Freedman (BF)
bound \cite{Freedman}. In this work, we shall set $m^{2}=-2$. Hence $\Delta_{-}=1$ and $\Delta_{+}=2$, and both $\Psi_-$ and $\Psi_+$ can serve as the source, which correspond to
the standard quantization and the alternative quantization, respectively \cite{C.P}.
According to the holographic dictionary, we can obtain the vacuum expectation of the dual operators on the boundary as
\begin{equation}
\left\langle j^{\mu}\right\rangle  =\frac{\delta S_{onshell}}{\delta a_{\mu}}=b^{\mu},\quad 
\left\langle O_{\pm}\right\rangle  =\frac{\delta S_{onshell}}{\delta\Psi_{\mp}}=\pm\Psi_{\pm}^{*}.\label{eq:order parameter dictionary}
\end{equation}
Here, $j^{\mu}$ is the $U(1)$ conserved current on the boundary. In particular,
$j^{t}$ is the conserved particle number density conjugating to the chemical
potential, $a_{t}=A_{t}|_{z=0}=\mu$. The vacuum expectation value of
the scalar operator $\left\langle O_{\pm}\right\rangle $ is interpreted
as the superfluids condensate in the holographic superfluids model.
If the condensate remains non-vanishing when the source is turned off, which indicates spontaneous breaking of the $U(1)$
symmetry, resulting in the boundary system being in a superfluid phase; otherwise, it remains in a normal phase. In the case of a uniform holographic superfluid in equilibrium, it has been shown in the reference \cite{Guo} that the system will transition to a superfluid phase when the chemical potential exceeds certain critical values. These values depend on different quantization schemes mentioned above.

In the following section, we focus on the superfluid phase while working with the soliton train configuration, which exhibits heterogeneity along the $x$ direction. Consequently, the soliton train configuration is inhomogeneous along the (x)-axis.
To effectively address the equations of motion mentioned earlier, it is necessary to choose the radial gauge $A_{z}=0$ for the U(1) gauge fields. For
simplicity, let us assume that the non-vanishing bulk fields are $\Psi:=z\psi$,
$A_{t}$ and $A_{x}$, which do not depend on the coordinate $\xi$. As a result, the equations of motion become
\begin{align}
0= & \partial_{t}^{2}\psi+\left(z+A_{x}^{2}-A_{t}^{2}+i\partial_{x}A_{x}-i\partial_{t}A_{t}\right)\psi+2iA_{x}\partial_{x}\psi\nonumber\\
 &-2iA_{t}\partial_{t}\psi-\partial_{x}^{2}\psi+3z^{2}\partial_{z}\psi+\left(z^{3}-1\right)\partial_{z}^{2}\psi,\label{eq:psi equation}\\
0= & \partial_{t}^{2}A_{x}-\partial_{t}\partial_{x}A_{t}-i\left(\psi\partial_{x}\psi^{*}-\psi^{*}\partial_{x}\psi\right)+2A_{x}\psi\psi^{*}\nonumber \\
&+3z^{2}\partial_{z}A_{x}+\left(z^{3}-1\right)\partial_{z}^{2}A_{x},\label{eq:Ax equation}\\
0= & \left(z^{3}-1\right)\partial_{z}^{2}A_{t}+3z^{2}\partial_{z}A_{t}-\partial_{x}^{2}A_{t}+\partial_{t}\partial_{x}A_{x}+2A_{t}\psi\psi^{*}\nonumber \\
 & +i\left(\psi^{*}\partial_{t}\psi-\psi\partial_{t}\psi^{*}\right),\label{eq:At equation}\\
0= & \partial_{t}\partial_{z}A_{t}+i\left(\psi\partial_{z}\psi^{*}-\psi^{*}\partial_{z}\psi\right)-\partial_{z}\partial_{x}A_{x},\label{eq:boundary current}
\end{align}
where the third one can be interpreted as the constraint equation.

\section{Static configuration of soliton train}

At the core of the black soliton, there is a density depletion, with the order parameter switching signs at the interface. Consequently, for static black solitons, the matter fields depend on both $x$ and $z$. Namely, $\psi=\psi\left(z,x\right)$, $A_{t}=A_{t}\left(z,x\right)$
and $A_{x}=A_{x}\left(z,x\right)$. We further recast the complex
scalar field $\psi$ in the form $\psi\left(z,x\right)=\phi\left(z,x\right)\exp\left[i\varphi\left(z,x\right)\right]$.
Then we substitute $\psi$, $A_{t}$, $A_{x}$ into equations $\left(\ref{eq:psi equation}\right)\sim
\left(\ref{eq:boundary current}\right)$
and obtain the following equations,
\begin{align}
0= & \left(z+A_{x}^{2}-A_{t}^{2}\right)\phi-2A_{x}\phi\partial_{x}\varphi-\partial_{x}^{2}\phi-\phi\left(\partial_{x}\varphi\right)^{2}\nonumber \\ &+3z^{2}\partial_{z}\phi+\left(z^{3}-1\right)\left[\partial_{z}^{2}\phi-\phi\left(\partial_{z}\varphi\right)^{2}\right],\label{eq:real part}\\
0= & \phi\partial_{x}A_{x}+2A_{x}\partial_{x}\phi+2\left(\partial_{x}\phi\right)\partial_{x}\varphi+\phi\partial_{x}^{2}\varphi+3z^{2}\phi\partial_{z}\varphi\nonumber \\
 & +\left(z^{3}-1\right)\left[2\left(\partial_{z}\phi\right)\partial_{z}\varphi+\phi\partial_{z}^{2}\varphi\right],\label{eq:imaginary part}\\
0= & -2\phi^{2}\partial_{x}\varphi+2A_{x}\phi^{2}+3z^{2}\partial_{z}A_{x}+\left(z^{3}-1\right)\partial_{z}^{2}A_{x},\label{eq:Ax}\\
0= & \left(z^{3}-1\right)\partial_{z}^{2}A_{t}+3z^{2}\partial_{z}A_{t}-\partial_{x}^{2}A_{t}+2\phi^{2}A_{t},\label{eq:At}\\
0= & 2\phi^{2}\partial_{z}\varphi-\partial_{z}\partial_{x}A_{x}.\label{eq:Ax-1}
\end{align}

For static soliton train solutions, the U(1) current in the x direction (the direction of the soliton train) is zero. Additionally, the current in the gravitational system should also be zero, namely, $j^{x}=j^{z}=0$, so we have
\begin{align}
\partial_{z}\varphi & =0\label{eq:cphi condition}\\
\partial_{x}\varphi & =A_{x}\label{eq:cphi Ax}
\end{align}
Given the constraint conditions above, (\ref{eq:Ax-1}) and (\ref{eq:Ax})
are satisfied automatically. To fix the local gauge symmetry U (1) in the x-direction for Schwarzschild coordinates, we set $A_{x}=0$, resulting in $\varphi$ being constant.
As a result, the equations of motion $\left(\ref{eq:real part}\right) \sim \left(\ref{eq:Ax-1}\right)$
will be simplified as

\begin{align}
0 & =(z-A_{t}^{2})\psi-\partial_{x}^{2}\psi+3z^{2}\partial_{z}\psi+(z^{3}-1)\partial_{z}^{2}\psi\label{eq:phi}\\
0 & =(z^{3}-1)\partial_{z}^{2}A_{t}+3z^{2}\partial_{z}A_{t}-\partial_{x}^{2}A_{t}+2\psi^{2}A_{t}\label{eq:At-1}
\end{align}
In the following, we shall numerically solve these nonlinear differential
equations using the pseudo-spectral method \cite{Guo2} in conjunction with Newton-Raphson iteration technique.

Ideally, the system should be infinite along the $x$ direction. However, for numerical calculations, a cutoff is necessary. We consider a box of size $1\times L$ in the $z$ and $x$ directions, respectively. In addition, two types of boundary conditions can be applied to these equations: standard quantization and alternative quantization. For the standard quantization, we designate $\Psi_{-}$ as the source. The boundary conditions are $\left.\psi\right|_{z=0}=0$
and $\left.A_{t}\right|_{z=0}=\mu$, where $\mu$ represents the chemical potential. Consequently, the superfluid order parameter is $\left\langle O\right\rangle |_{z=0}=\partial_{z}\psi$. For alternative quantization, the source
$\Psi_{+}$ is considered. The boundary condition at $z=0$ is $\partial_{z}\psi=0$
and $A_{t}=\mu$. As a result, the superfluid order parameter is
$\left\langle O\right\rangle |_{z=0}=\psi$. In addition, a regular
condition must be imposed at the $z=1$ boundary. We apply the Chebyshev pseudo-spectral method \cite{Guo2} to discretize space in the $z$ direction.
To achieve a periodic soliton train configuration, it is necessary to apply periodic boundary conditions in the $x$ direction using Fourier pseudo-spectral method. Taking into account these boundary conditions, we employ the Newton iteration method to solve the equations of motion.

\begin{figure}
\includegraphics[scale=0.3]{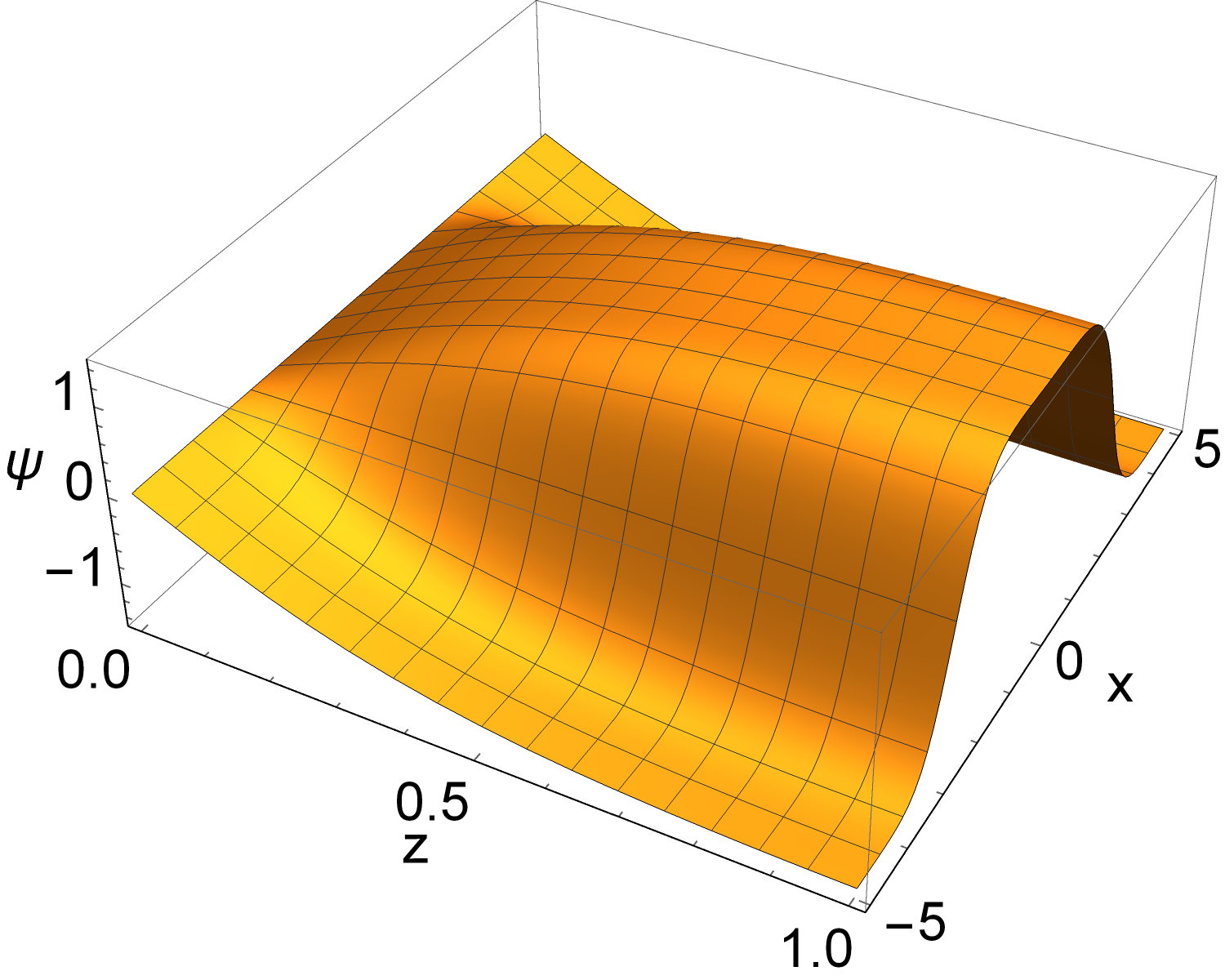}
\includegraphics[scale=0.3]{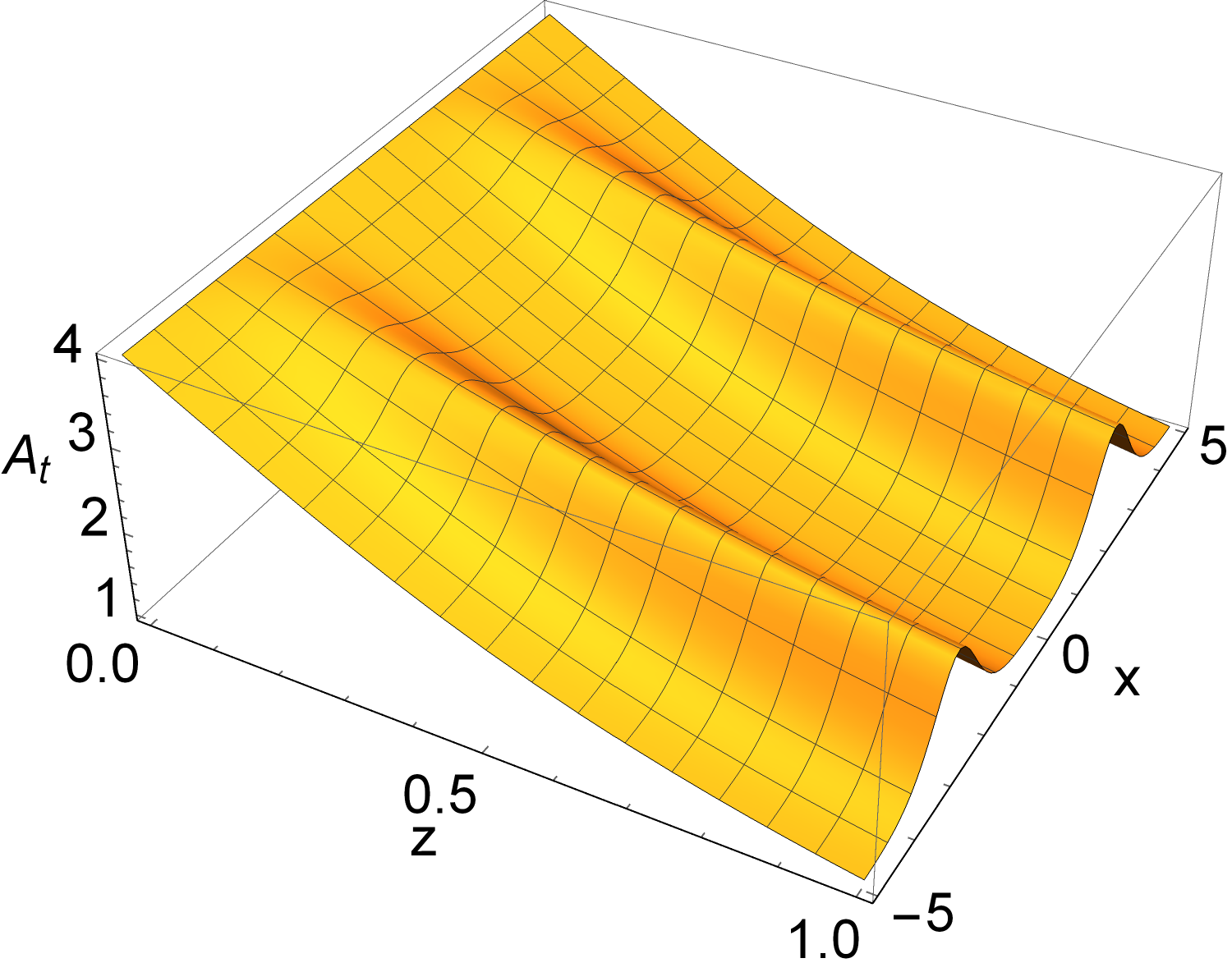}

\caption{The bulk configurations of the fields $\psi\left(z,x\right)$ (upper) and $A_{t}\left(z,x\right)$ (lower) for $\mu=4,L=10$
at standard quantization scheme\label{standard field for black soliton trains}}
\end{figure}

\begin{figure}
\includegraphics[scale=0.3]{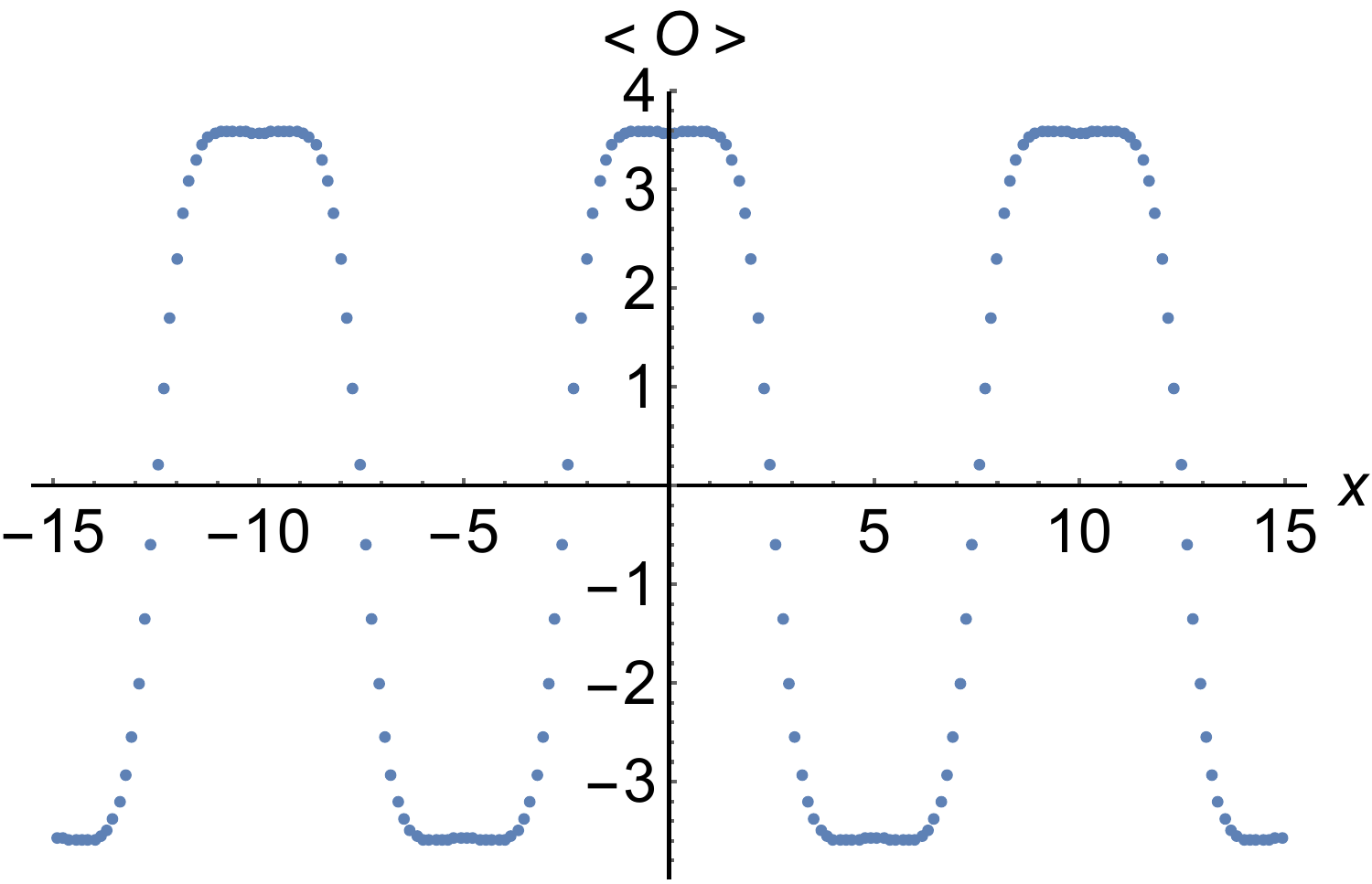}
\includegraphics[scale=0.3]{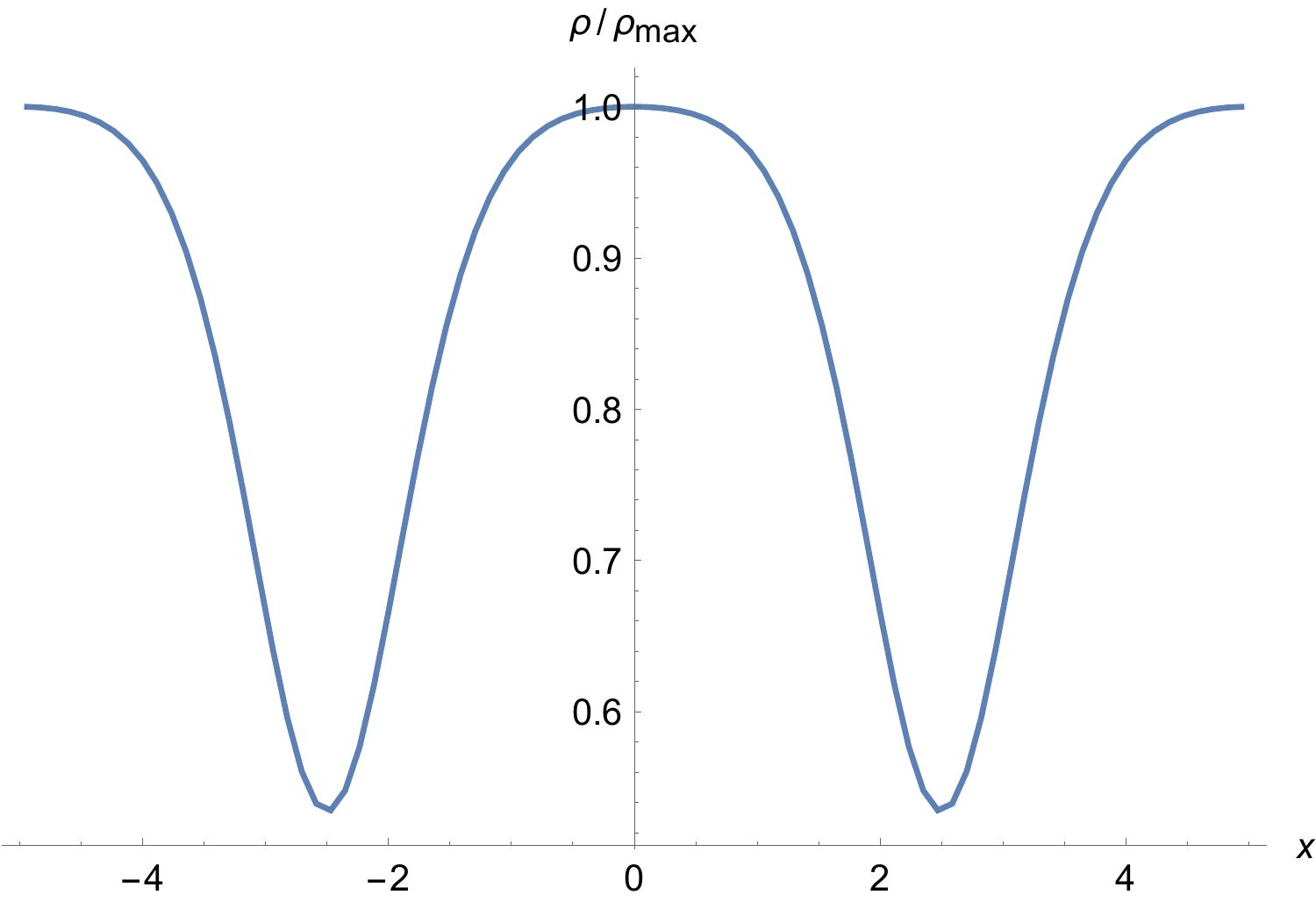}

\caption{The order parameter (upper) and normalized particle number density
(lower) for standard quantization with $\mu=4,L=10$\label{order and density in standard black}}
\end{figure}

\begin{figure}
\includegraphics[scale=0.3]{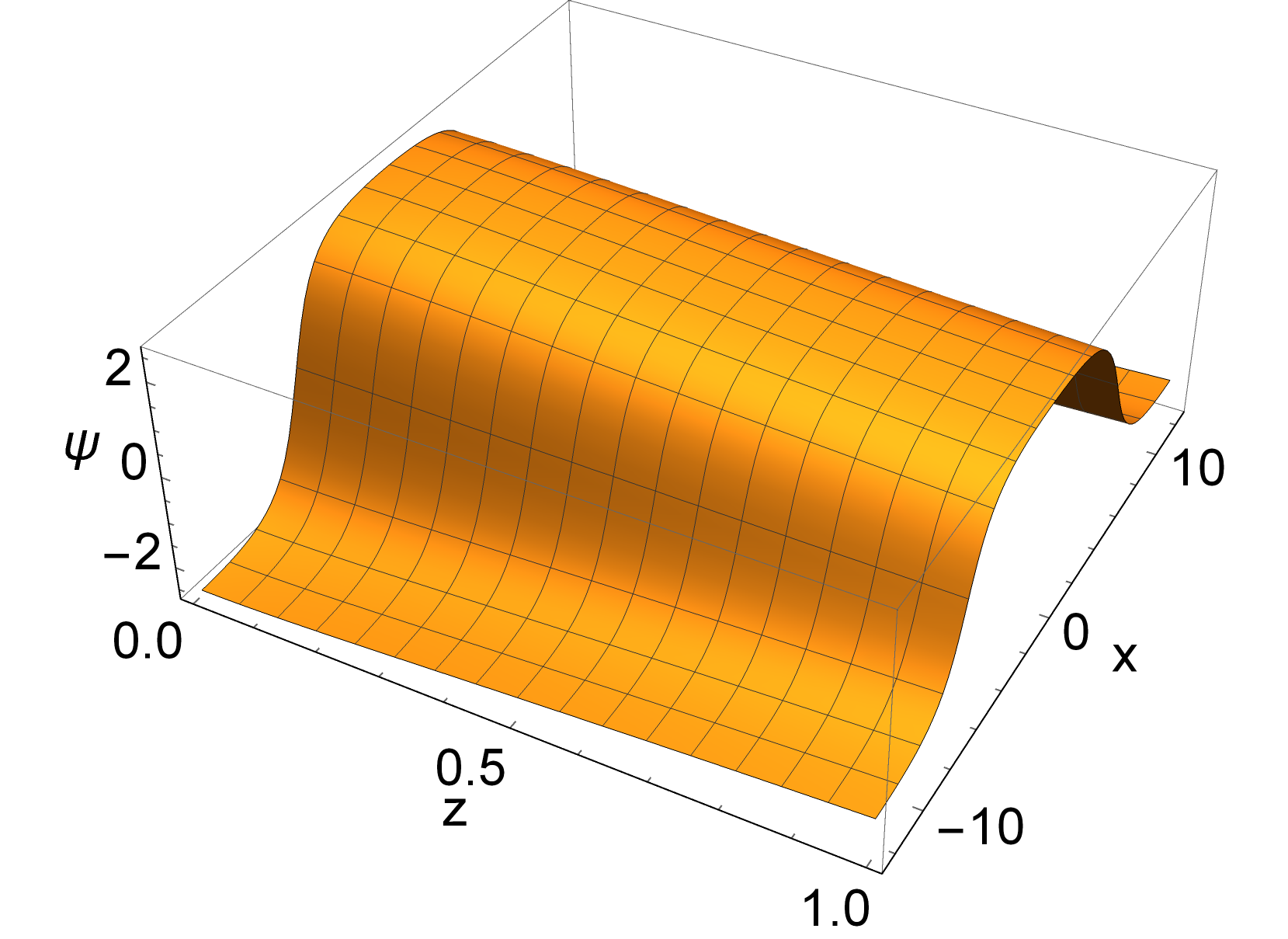}
\includegraphics[scale=0.3]{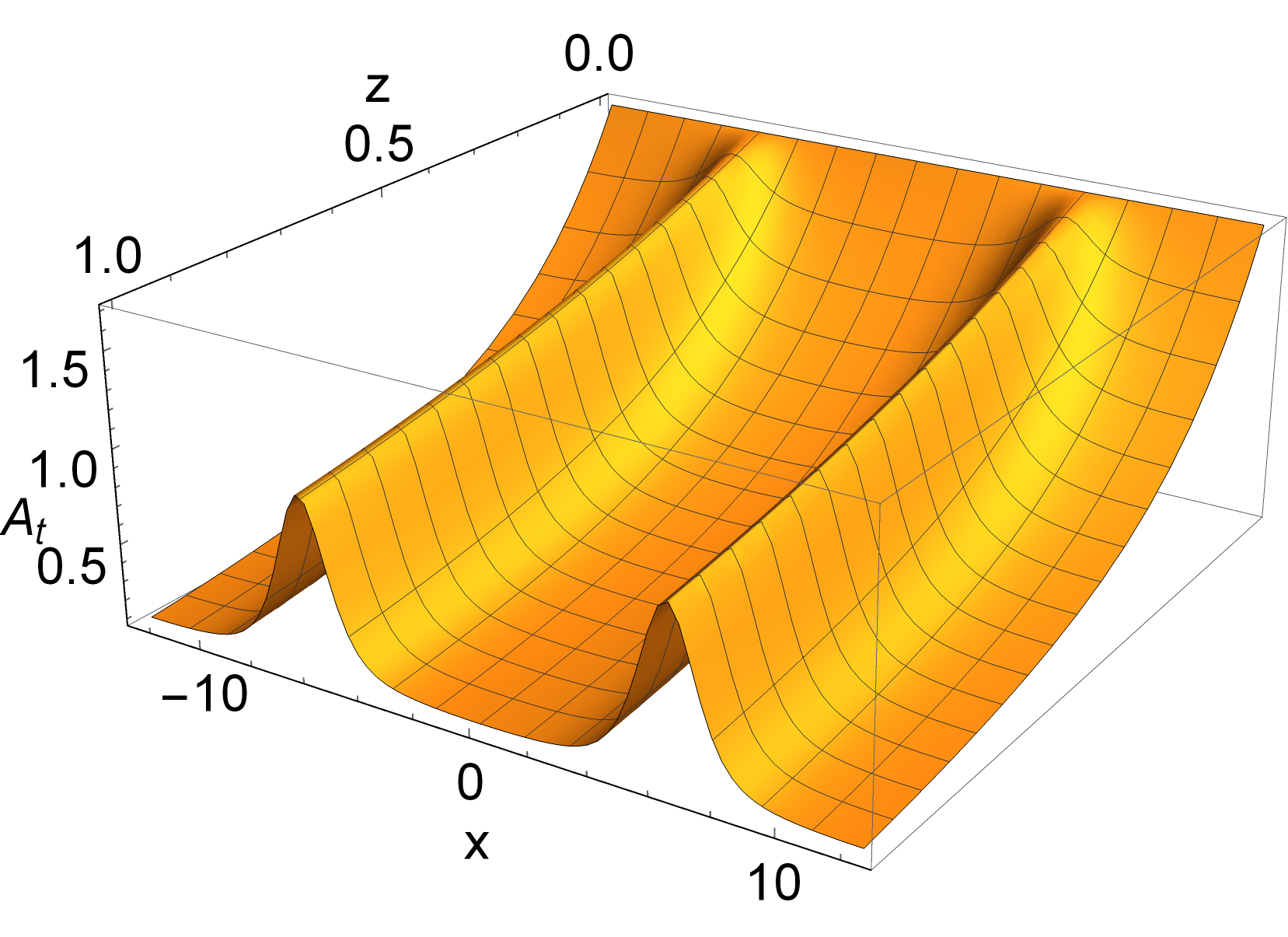}

\caption{The bulk configurations of the fields $\psi\left(z,x\right)$ (upper) and $A_{t}\left(z,x\right)$ (lower) for alternative
quantization scheme with $\mu=1.7,L=8\pi$\label{fig:alternative-bulk-configurations}}

\end{figure}

\begin{figure}
\includegraphics[scale=0.3]{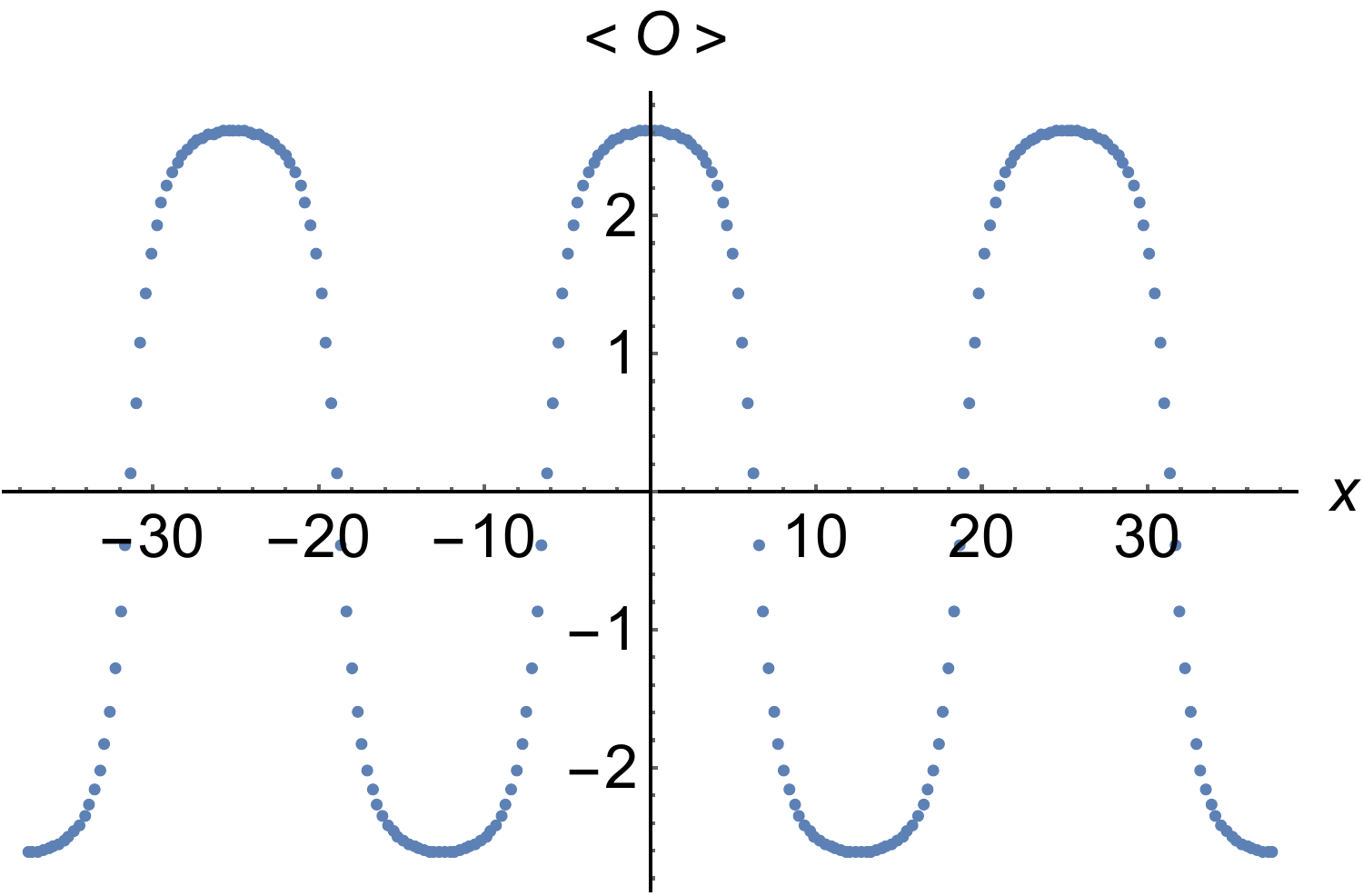}
\includegraphics[scale=0.3]{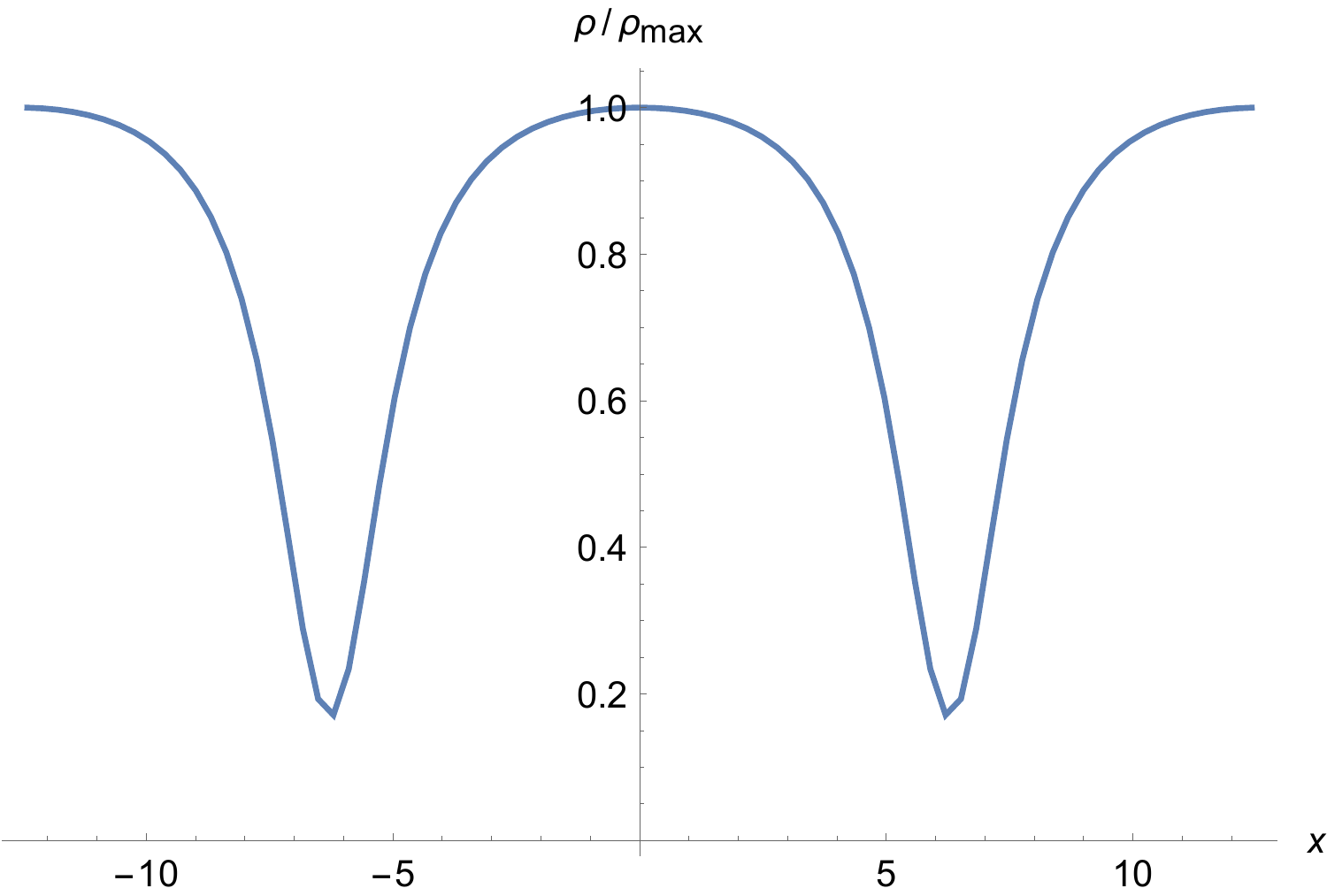}

\caption{The order parameter (upper) and normalized particle number density
(lower) for alternative quantization with $\mu=1.7,L=8\pi$\label{fig:alternative The-order-parameter and density} }

\end{figure}

In standard quantization, we illustrate the numerical results of the bulk field configurations for the soliton train at the chemical potential of $\mu=4$ in Fig.\ref{standard field for black soliton trains}. Fig.\ref{order and density in standard black} shows the order parameter and normalized particle number density as a function of $x$, with the lattice spacing $L=10$. On the AdS boundary, the asymptotic behavior of $A_{t}$ is given by $A_{t}=\mu-\rho z$. Therefore, the particle number density can be obtained as $\rho=-\partial_{z}A_{t}|_{z=0}$.
We present numerical results of the bulk field configurations for the soliton train in Fig.\ref{fig:alternative-bulk-configurations} under alternative quantization, with a chemical potential of $\mu=1.7$ and a lattice spacing of $L=8\pi$. The corresponding order parameter and the normalized particle
number density are depicted in Fig.\ref{fig:alternative The-order-parameter and density}. Comparing Fig.\ref{order and density in standard black} with Fig.\ref{fig:alternative The-order-parameter and density}, we observe a greater depletion of charge density at the soliton core for alternative quantization compared to standard quantization. This observation aligns with findings in reference \cite{gaomeng} and suggests that the two quantization schemes correspond to BCS superfluid and BEC superfluid states, respectively \cite{gaomeng}.

\section{linear dynamics evolution of soliton train}

In this section, we discuss the linear analysis of soliton train configuration via Quasi-normal Modes (QNMs) technique. To calculate the QNMs, it is recommended to reformulate the equations of motion $\left(\ref{eq:psi equation}\right)\sim\left(\ref{eq:boundary current}\right)$
within the Hamilton formalism, as suggested by \cite{Guo}, rather than employing the Eddington-Finkelstein coordinates,
\begin{align}
0= & \partial_{t}\psi-iA_{t}\psi-P,\label{eq:P}\\
0= & \partial_{t}A_{x}-\partial_{x}A_{t}-\Pi_{x},\label{eq:Pi}\\
0= & -\partial_{t}P+iA_{t}P-\left(z+i\partial_{x}A_{x}+A_{x}^{2}\right)\psi-3z^{2}\partial_{z}\psi\nonumber \\
 &+\left(1-z^{3}\right)\partial_{z}^{2}\psi+\partial_{x}^{2}\psi-2iA_{x}\partial_{x}\psi,\label{eq:Pt-1}\\
0= & \partial_{t}\Pi_{x}-i\left(\psi\partial_{x}\psi^{*}-\psi^{*}\partial_{x}\psi\right)+2A_{x}\left|\psi\right|^{2}+3z^{2}\partial_{z}A_{x}\nonumber \\
&+\left(z^{3}-1\right)\partial_{z}^{2}A_{x},\label{eq:Pit}\\
0= & 3z^{2}\partial_{z}A_{t}+\left(z^{3}-1\right)\partial_{z}^{2}A_{t}+\partial_{x}\Pi_{x}+i\left(P\psi^{*}-\psi P^{*}\right),\label{eq:constrain}\\
0= & \left(1-z^{3}\right)\partial_{t}\partial_{z}A_{t}+i\left(1-z^{3}\right)\left(\psi\partial_{z}\psi^{*}-\psi^{*}\partial_{z}\psi\right)\nonumber \\
 &-\left(1-z^{3}\right)\partial_{z}\partial_{x}A_{x}.\label{eq:throw away}
\end{align}

Where, $P$ represents the conjugate momentum of $\psi$, $\Pi_{x}$ represents the
conjugate momentum of $A_{x}$. Similarly, we will designate (\ref{eq:constrain})
as the constraint equation. Furthermore, the corresponding linear perturbation
equations are as follows.

\begin{align}
\partial_{t}\left(\delta\psi\right)= & \delta P+i\left(\psi\delta A_{t}+A_{t}\delta\psi\right)\label{eq:deltaP}\\
\partial_{t}\left(\delta A_{x}\right)= & \delta\Pi_{x}+\partial_{x}\left(\delta A_{t}\right)\label{eq:deltaPi}\\
\partial_{t}\delta P= & i\left(P\delta A_{t}+A_{t}\delta P\right)-\left[i\partial_{x}\left(\delta A_{x}\right)+2A_{x}\delta A_{x}\right]\psi\nonumber \\
 &-\left(i\partial_{x}A_{x}+A_{x}^{2}\right)\delta\psi-z\delta\psi-3z^{2}\left(\partial_{z}\delta\psi\right)\nonumber \\
 &+\left(1-z^{3}\right)\partial_{z}^{2}\left(\delta\psi\right)+\partial_{x}^{2}\left(\delta\psi\right)\nonumber \\
 &-2i\left(\delta A_{x}\right)\partial_{x}\psi-2iA_{x}\partial_{x}\left(\delta\psi\right)\label{eq:deltaPt}\\
\partial_{t}\Pi_{x}= & i\left[\delta\psi\partial_{x}\psi^{*}+\psi\partial_{x}\left(\delta\psi\right)^{*}-\left(\delta\psi\right)^{*}\partial_{x}\psi-\psi^{*}\partial_{x}\left(\delta\psi\right)\right]\nonumber \\
 & -2\left[\delta A_{x}\left|\psi\right|^{2}+A_{x}\psi^{*}\delta\psi+A_{x}\psi\left(\delta\psi\right)^{*}\right]\nonumber \\
 &-3z^{2}\partial_{z}\left(\delta A_{x}\right)+\left(1-z^{3}\right)\partial_{z}^{2}\left(\delta A_{x}\right)\label{eq:deltaPit}\\
0= & 3z^{2}\partial_{z}\delta A_{t}+\left(z^{3}-1\right)\partial_{z}^{2}\delta A_{t}+\partial_{x}\delta\Pi_{x}\nonumber \\
 &+i\left[\psi^{*}\delta P+P\left(\delta\psi\right)^{*}-\delta\psi P^{*}-\psi\left(\delta P\right)^{*}\right]\label{eq:yushu}\\
\partial_{t}\partial_{z}\delta A_{t}= & i\left[\left(\partial_{z}\psi\right)\left(\delta\psi\right)^{*}+\psi^{*}\partial_{z}\delta\psi-\left(\partial_{z}\psi^{*}\right)\delta\psi-\psi\partial_{z}\left(\delta\psi\right)^{*}\right]\nonumber \\
 &+\partial_{x}\partial_{z}\delta A_{x}\label{eq:shouheng-1}
\end{align}

Considering that there is no black hole in the AdS-soliton background, the bulk charge does not enter the black hole. Since the bulk charge is equivalent to the boundary charge, the conservation of boundary charge is ensured, following from the conservation mandated by the Klein-Gordon equation. Consequently, we can ignore the perturbation equation $\left(\ref{eq:shouheng-1}\right)$.

Since the soliton train is regularly spaced along the $x$ direction, we apply Bloch's theorem to assume the form of perturbation bulk fields as follows.

\begin{align}
\delta\psi= & p_{1}\left(z,x\right)e^{-i\omega t+ikx}+p_{2}^{*}\left(z,x\right)e^{i\omega^{*}t-ikx}\label{eq:deltapsi}\\
\delta At= & a\left(z,x\right)e^{-i\omega t+ikx}+a^{*}\left(z,x\right)e^{i\omega^{*}t-ikx}\label{eq:deltaAt}\\
\delta A_{x}= & b\left(z,x\right)e^{-i\omega t+ikx}+b^{*}\left(z,x\right)e^{i\omega^{*}t-ikx}\label{eq:deltaAx}\\
\delta P= & q_{1}\left(z,x\right)e^{-i\omega t+ikx}+q_{2}^{*}\left(z,x\right)e^{i\omega^{*}t-ikx}\label{eq:deltaPP}\\
\delta\Pi_{x}= & c\left(z,x\right)e^{-i\omega t+ikx}+c^{*}\left(z,x\right)e^{i\omega^{*}t-ikx}\label{eq:deltaPix}
\end{align}

Next, substitute the perturbation bulk fields into equations $\left(\ref{eq:deltaP}\right)\sim\left(\ref{eq:yushu}\right)$ to derive the linear perturbation equations,

\begin{align}
0= & ip_{1}A_{t}+q_{1}+ia\psi+i\omega p_{1}\label{eq:1}\\
0= & ia\psi^{*}+ip_{2}A_{t}-i\omega p_{2}-q_{2}\label{eq:2}\\
0= & -iak-i\omega b-c-\partial_{x}a\label{eq:3}\\
0= & a\psi^{*}A_{t}-iq_{2}A_{t}+2ib\partial_{x}\psi^{*}-kb\psi^{*}-k^{2}p_{2}-zp_{2}\nonumber \\
 &+i\omega q_{2}+\partial_{x}^{2}p_{2}+\left(1-z^{3}\right)\partial_{z}^{2}p_{2}+i\psi^{*}\partial_{x}b\nonumber \\ &+2ik\partial_{x}p_{2}-3z^{2}\partial_{z}p_{2}\label{eq:4}\\
0= & a\psi A_{t}+iq_{1}A_{t}-2ib\partial_{x}\psi+k\psi b+\partial_{x}^{2}p_{1}+\left(1-z^{3}\right)\partial_{z}^{2}p_{1}\nonumber \\
 &-i\psi\partial_{x}b+2ik\partial_{x}p_{1}-3z^{2}\partial_{z}p_{1}-k^{2}p_{1}-zp_{1}+i\omega q_{1}\label{eq:5}\\
0= & 2b\psi\psi^{*}-ic\omega-\left(1-z^{3}\right)\partial_{z}^{2}b+i\psi^{*}\partial_{x}p_{1}-i\psi\partial_{x}p_{2}\nonumber \\
 &-ip_{1}\partial_{x}\psi^{*}+ip_{2}\partial_{x}\psi+3z^{2}\partial_{z}b-kp_{1}\psi^{*}+kp_{2}\psi\label{eq:6}\\
0= & ick-\left(1-z^{3}\right)\partial_{z}^{2}a+\partial_{x}c+3z^{2}\partial_{z}a+A_{t}p_{1}\psi^{*}\nonumber \\
 &+\psi p_{2}A_{t}+iq_{1}\psi^{*}-iq_{2}\psi\label{eq:7}
\end{align}

In accordance with the background static solutions, we apply Dirichlet conditions for $p_{1},p_{2},a,b,c,q_{1},q_{2}$ at $z=0$ and periodic boundary conditions in the $x$ direction. In addition, regular conditions are imposed at $z=1$. By solving generalized eigenvalue problems, we determine the eigenvalue $\omega$ and eigenvector corresponding to a specified $k$ value.

\begin{figure}
\includegraphics[scale=0.32]{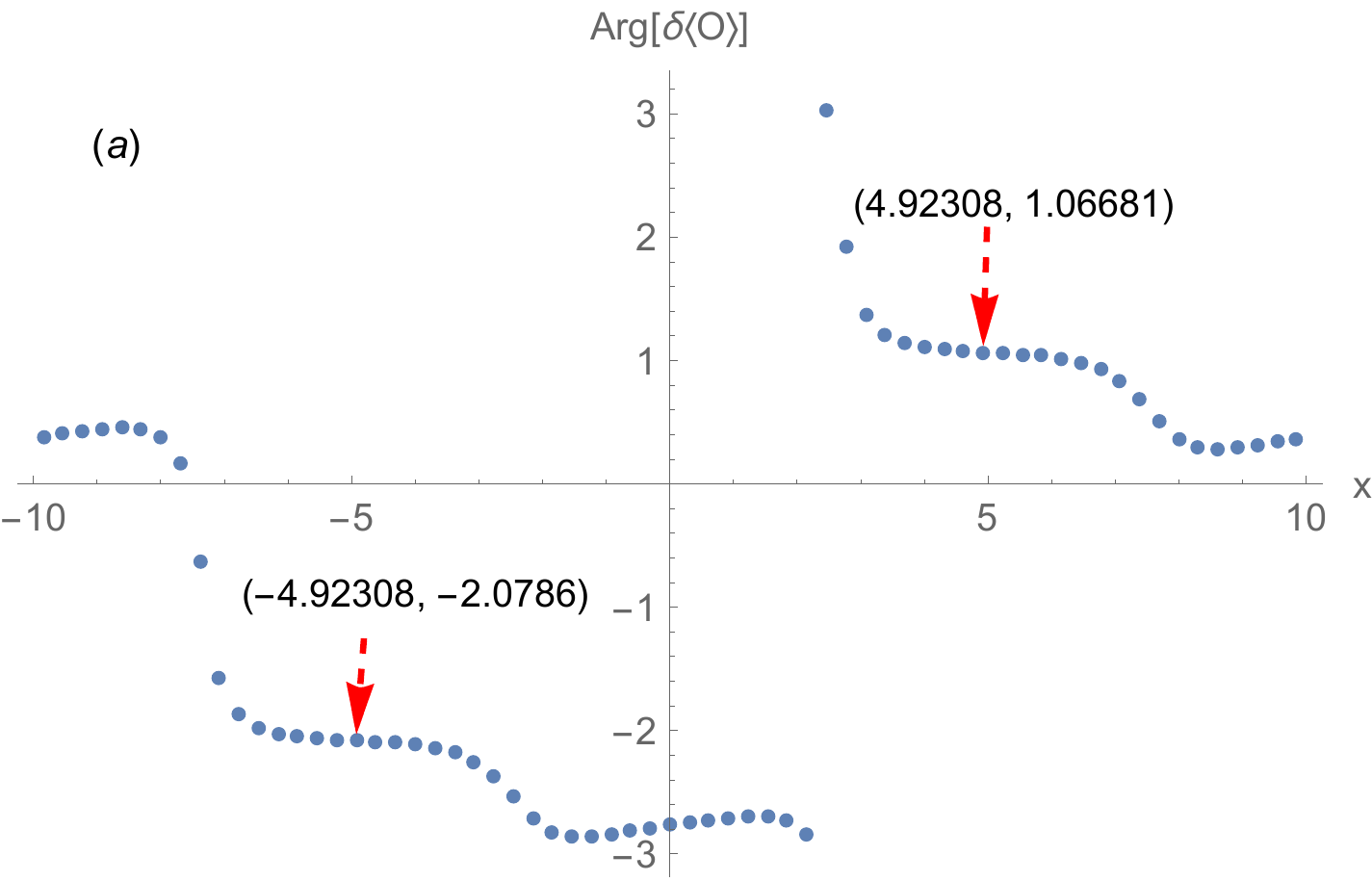}
\includegraphics[scale=0.32]{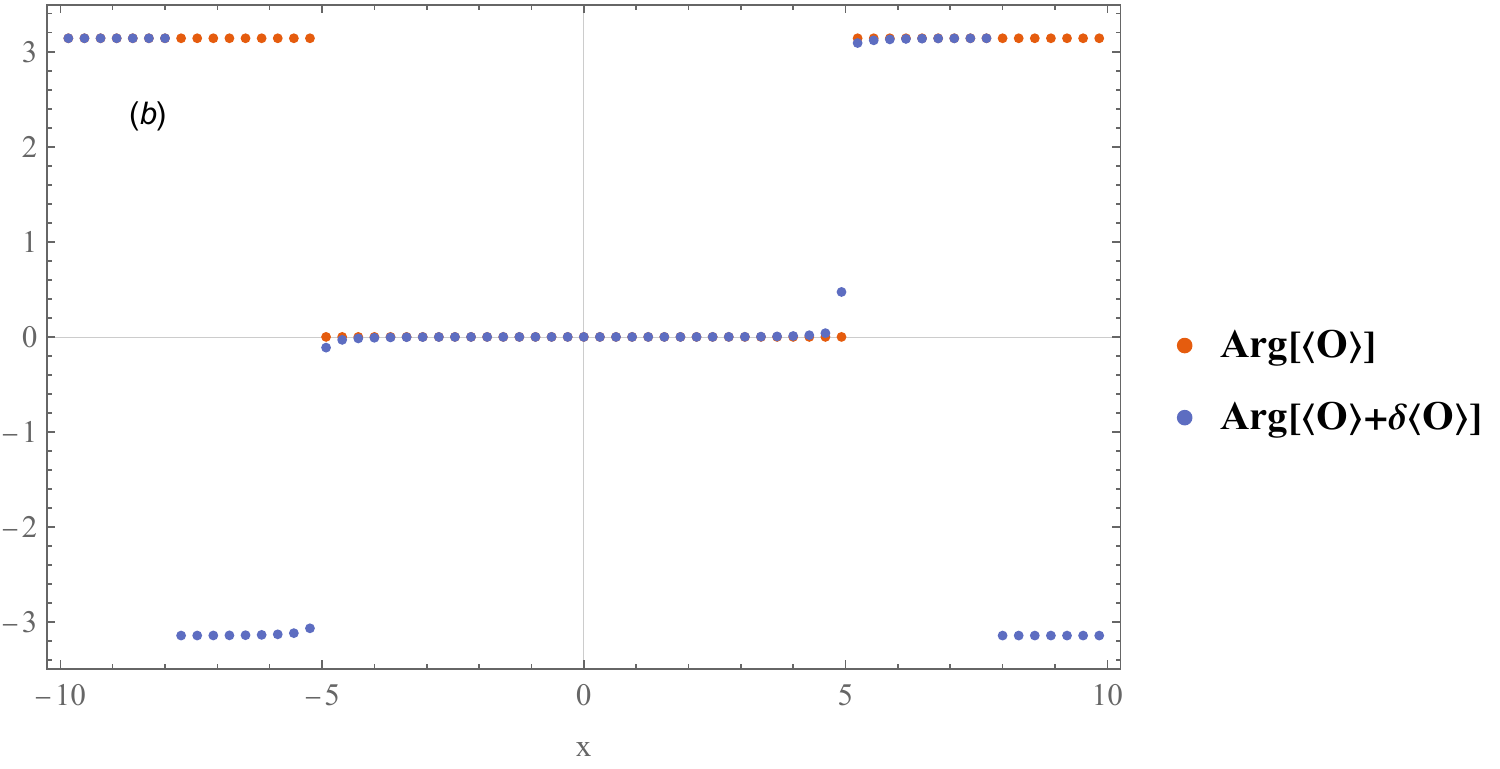}
\includegraphics[scale=0.32]{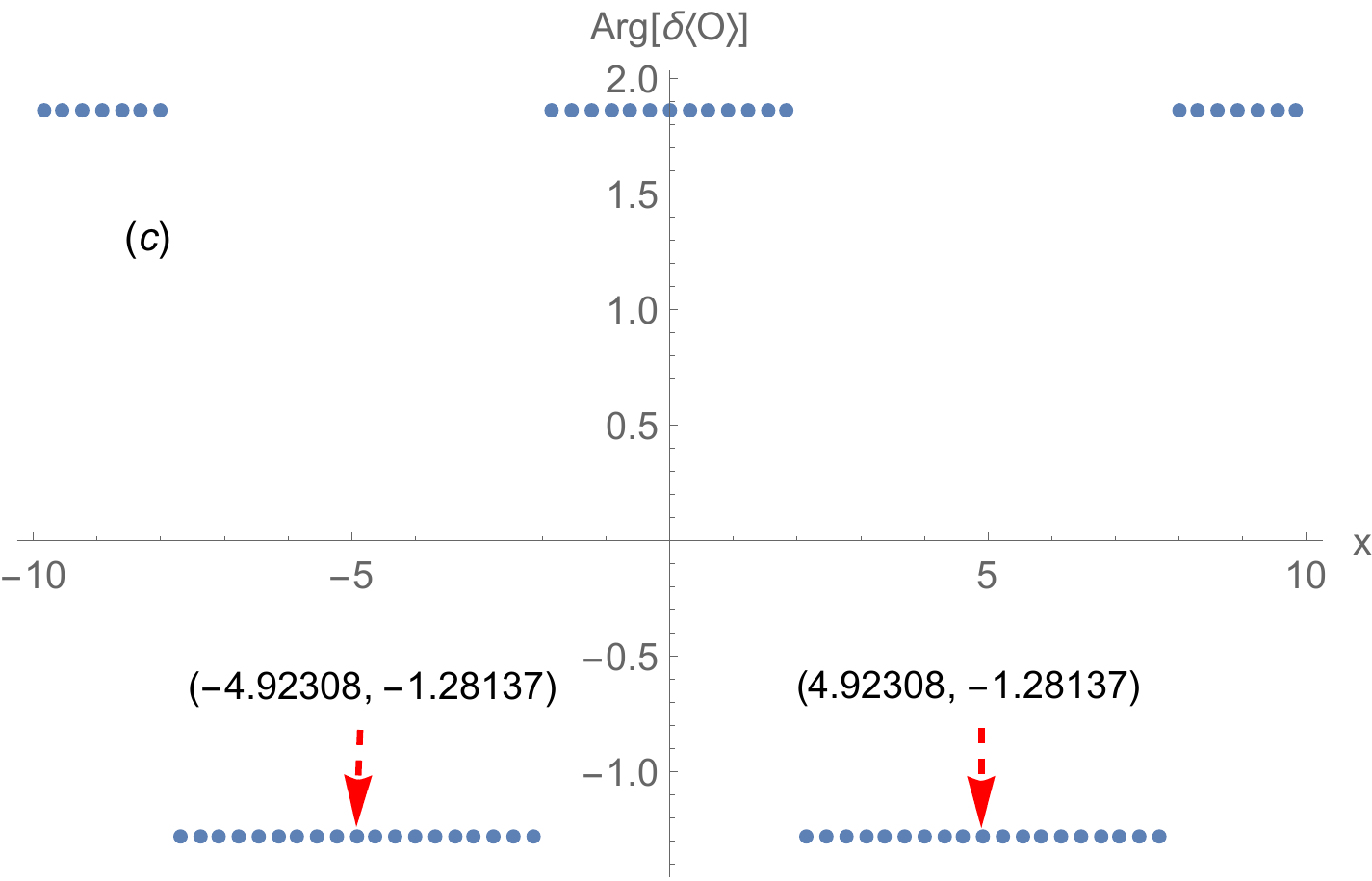}
\includegraphics[scale=0.32]{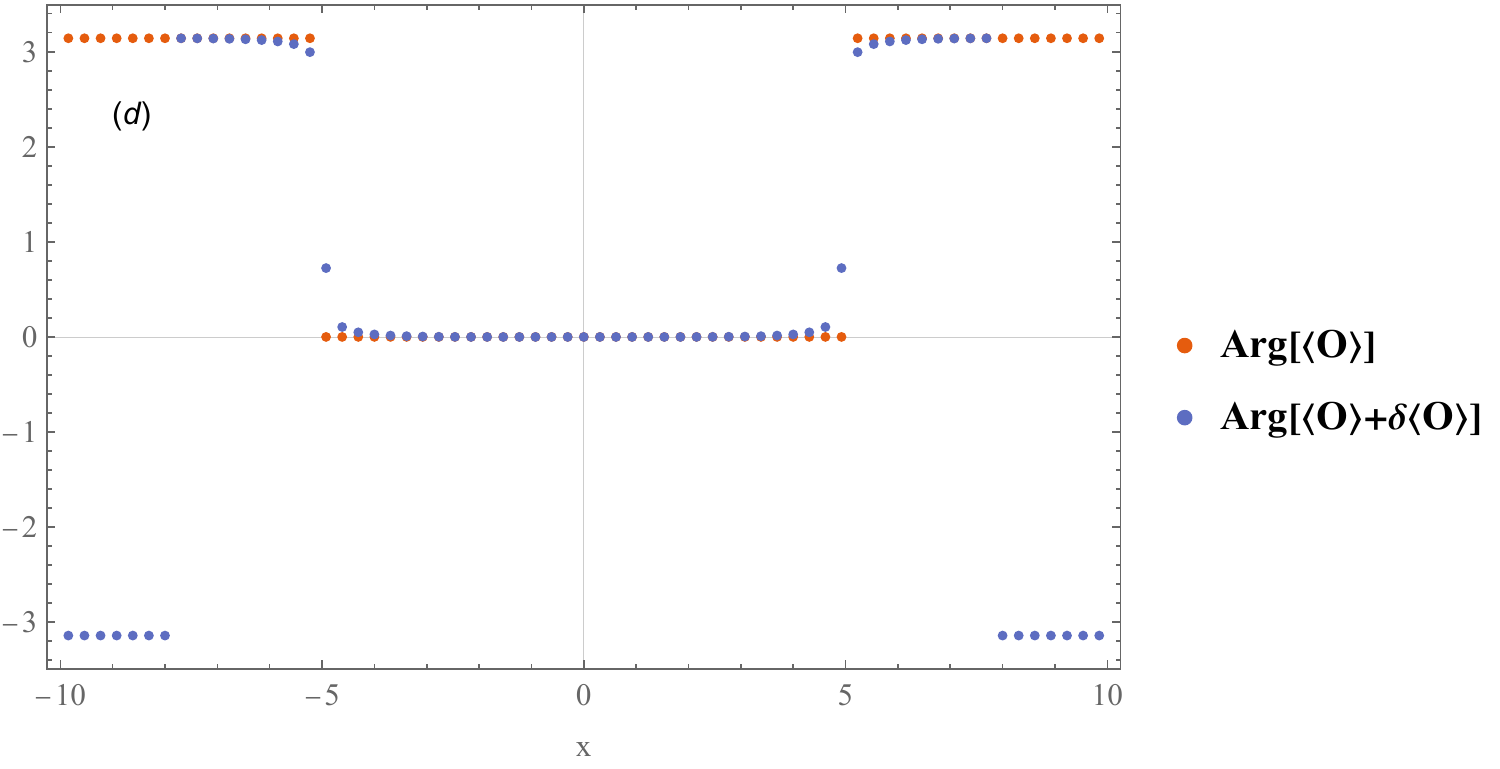}

\caption{The phase argument of the small elastic mode (upper) and large elastic
mode (lower), physical parameter is $\mu=2.3,L=20,k=0$. Since the
length scale is $L=20$, thus soliton core lies in $x=\pm5$. We can
identify from (a) and (c) that the phase difference at the two adjacent
soliton cores is around $3.14$ for small elastic mode, however it
is zero for large elastic mode. It shows, from (b) and (d), that two
elastic modes make different effect on soliton train from phase argument
aspect.\label{fig:The-phase-argument}}

\end{figure}

First of all, our analysis reveals the emergence of two elastic modes and one phonon mode upon perturbing the soliton train. One elastic mode, characterized by a larger eigenvalue, is termed the "large elastic mode", while the other, with a smaller eigenvalue, is referred to as the "small elastic mode". Initially, our goal is to distinguish between these two elastic modes and to figure out their distinct impacts on the soliton train configuration. To achieve clarity, we adopt three distinct perspectives to explore this issue. The phase distributions of the two elastic modes are illustrated in Fig. \ref{fig:The-phase-argument}.
Figs. \ref{fig:The-phase-argument}(a) and \ref{fig:The-phase-argument}(c)
depict the phase arguments of the small elastic mode and large elastic
mode, respectively. Notably, we observe a phase difference of $\pi$ at the two adjacent soliton cores ($x=\pm\frac{L}{4}$) for the small elastic
mode, while this difference is zero for the large elastic mode. In addition, we conducted calculations to assess the phase difference for both elastic modes under varying chemical potential values and length scale value (denoted by ($L$)), revealing consistent results.
Fig. \ref{fig:The-phase-argument}(b) and \ref{fig:The-phase-argument}(d)
describe the impact of the two types of elastic modes on the soliton train concerning phase arguments. Here, we discern that while the small elastic mode alters the phase difference between adjacent soliton cores, the large elastic mode does not induce such alterations. Consequently, we can qualitatively distinguish between these two elastic modes based on their distinct influences. The small elastic mode induces phase deformation in the cores of adjacent soliton trains, whereas the large elastic mode does not. Interestingly, the large mode is gapped, while the small mode remains gapless.

\begin{figure}
\includegraphics[scale=0.32]{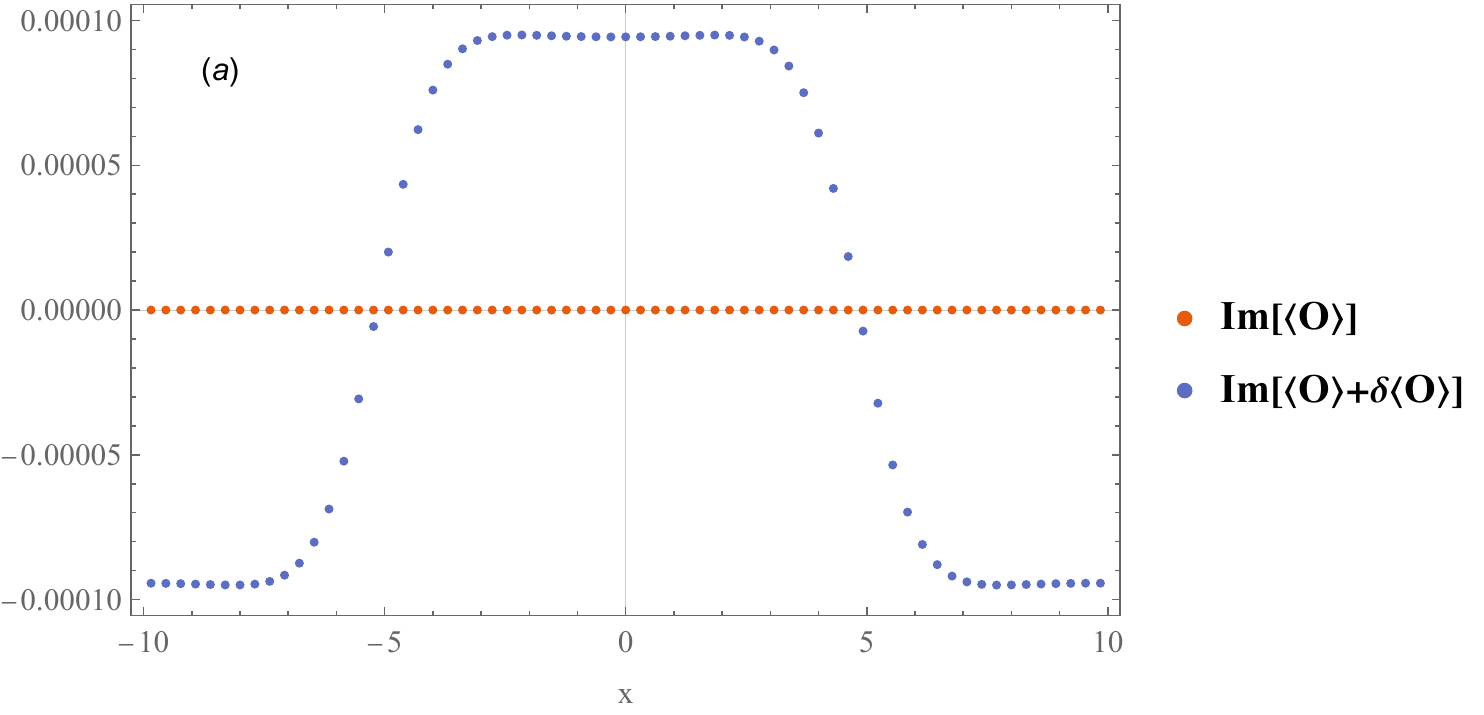}
\includegraphics[scale=0.32]{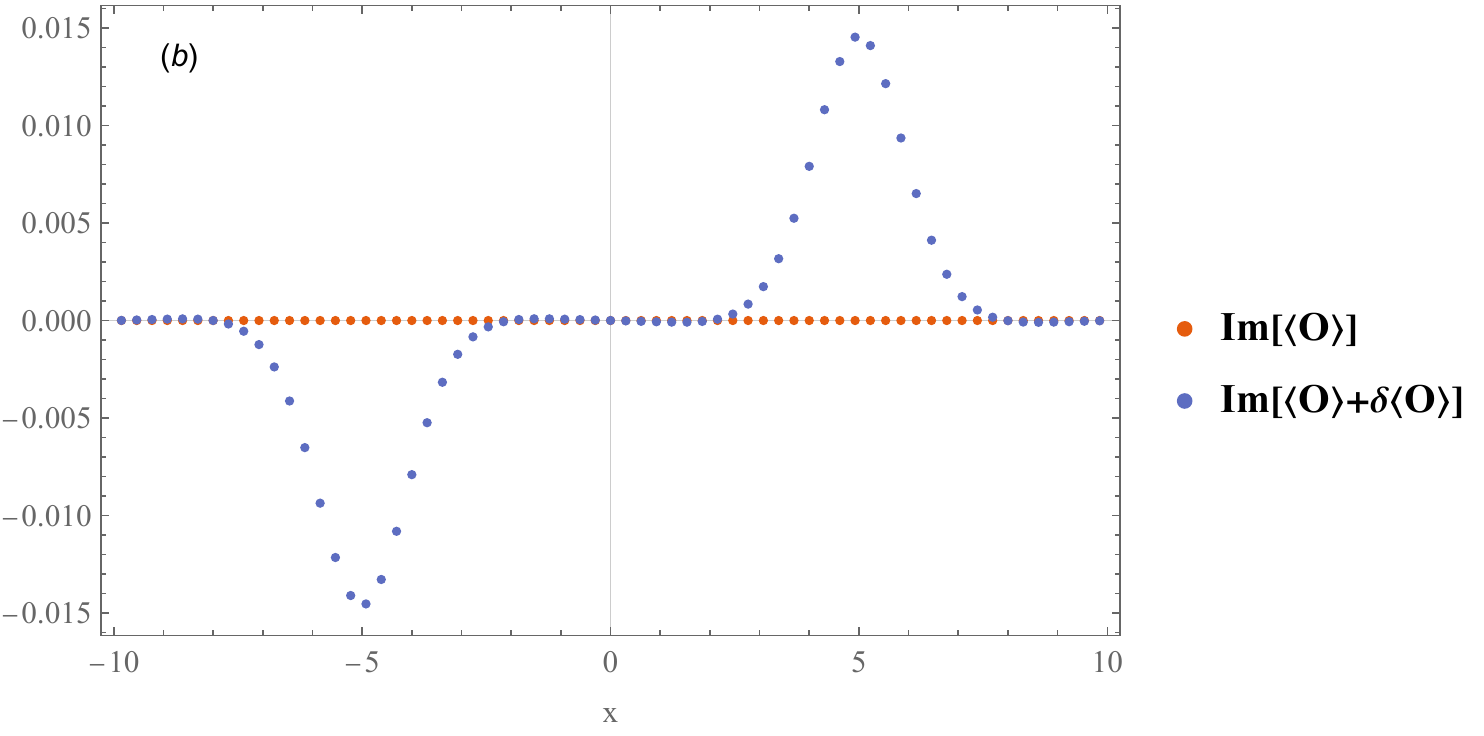}
\includegraphics[scale=0.32]{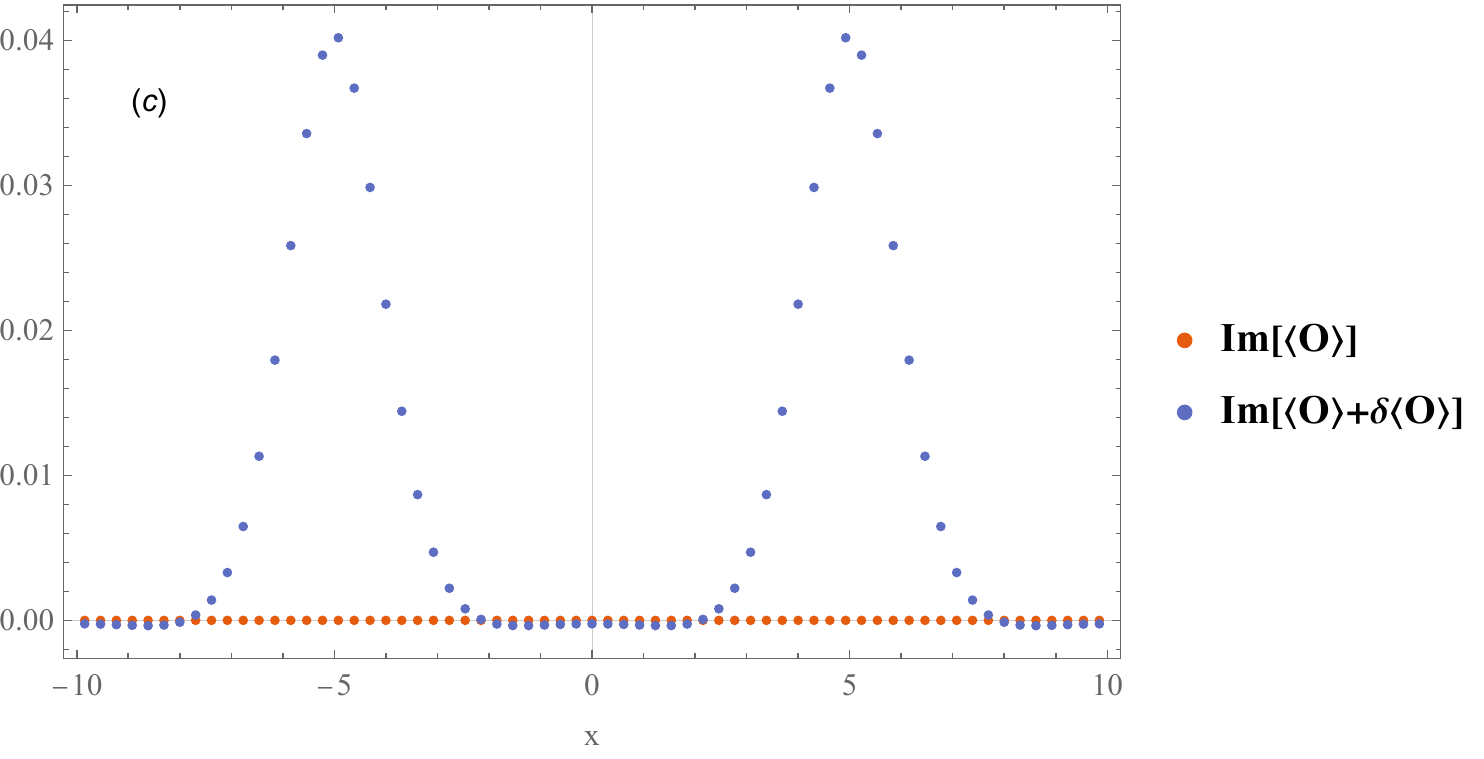}

\caption{Effect of three kinds of modes on soliton train at imaginary part
aspect, where $\mu=2.3,L=20,k=0$. (a): phonon mode. (b): small elastic
mode. (c): large elastic mode.\label{fig:imaginary part effect}}

\end{figure}

In the following, we will make a comparison of the impact of the imaginary components of the three kinds of modes (one phonon mode and two
elastic modes) on the soliton train, as depicted in Fig.\ref{fig:imaginary part effect}.
Fig. \ref{fig:imaginary part effect}(a) describes the phonon mode situation.
We observe that phonon mode does not affect imaginary part of
soliton cores. Specifically, the imaginary part at the cores of the soliton train remains zero even after imposing perturbations of the phonon mode. Fig. \ref{fig:imaginary part effect}(b)
shows that a small elastic mode significantly impacts the soliton train, as evidenced by the observation that the imaginary part at the soliton cores deviates from zero. Remarkably,
imaginary component of soliton cores exhibits oscillatory behavior within the complex plane. Here, we present a depiction of this oscillation over a single period, referring to it as "grayness oscillation" \cite{key-16}. Consequently, we observe that perturbations from the small elastic mode can induce a shift of the soliton train from the real plane to the complex plane.
Fig. \ref{fig:imaginary part effect}(c) demonstrates the large mode situation
and also changes the imaginary part at two soliton cores, but their variations are the
same.

\begin{figure}
\includegraphics[scale=0.26]{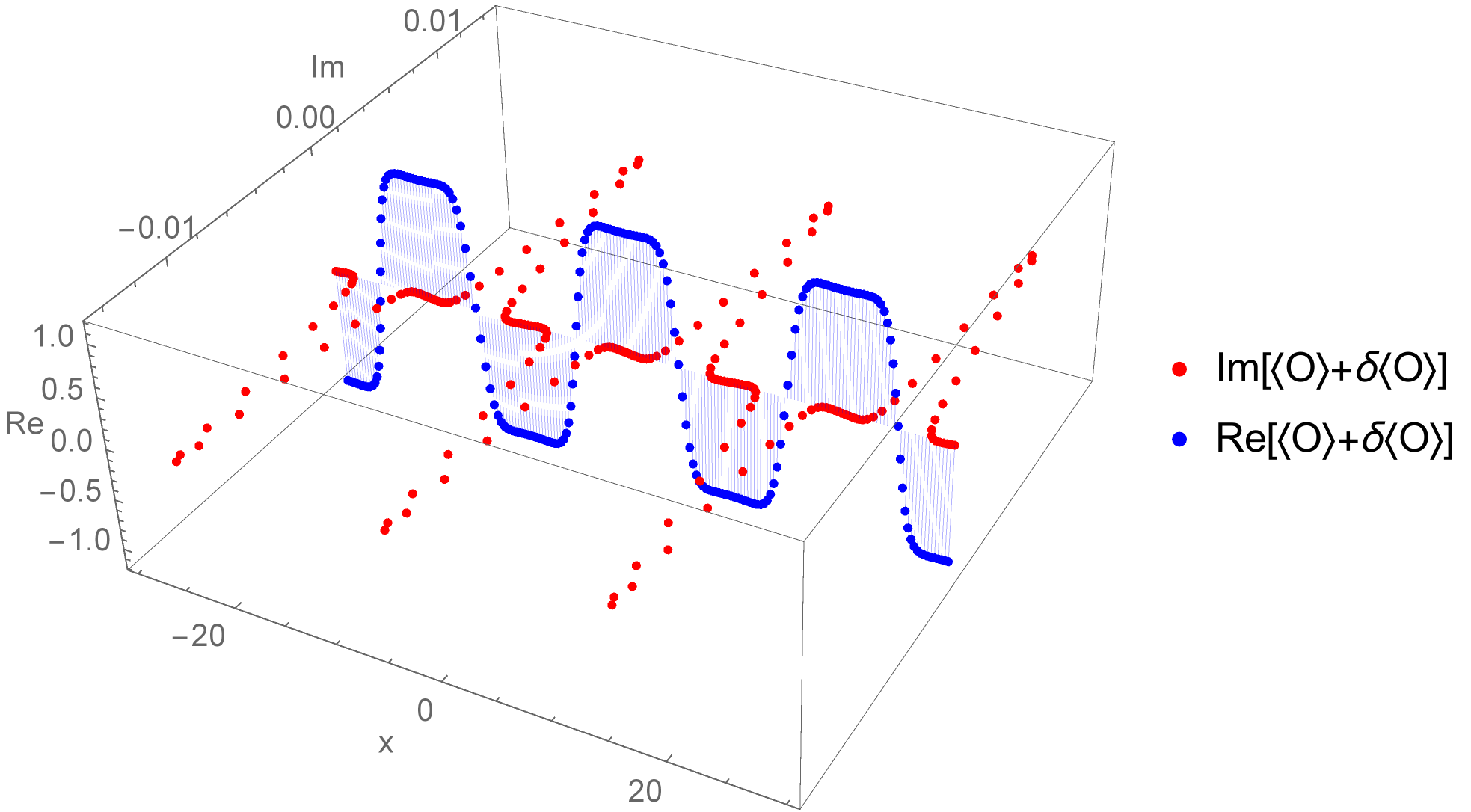}

\caption{Real part and imaginary part together for small elastic mode, where
the parameters are $\mu=2.3,L=20,k=0$.\label{fig:Real and imaginary} }

\end{figure}

To display the feature of grayness oscillation, we draw both real part
and imaginary part together in Fig. \ref{fig:Real and imaginary}.
By examining this visualization, we can identify the transition of the soliton train from the real plane to the complex plane under the influence of small elastic mode perturbation, manifesting as a distinct grayness oscillation.

\begin{figure}
\includegraphics[scale=0.32]{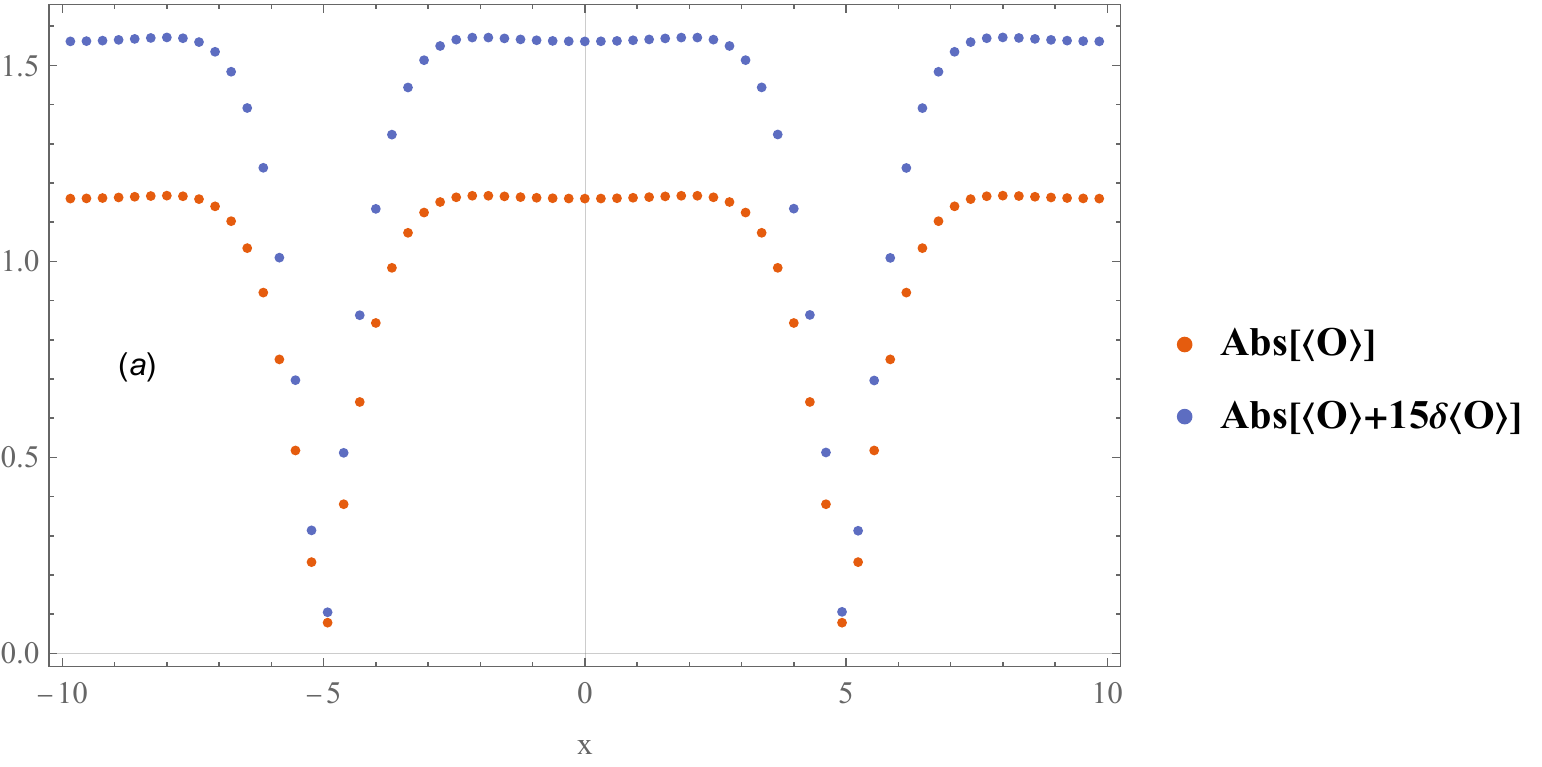}
\includegraphics[scale=0.32]{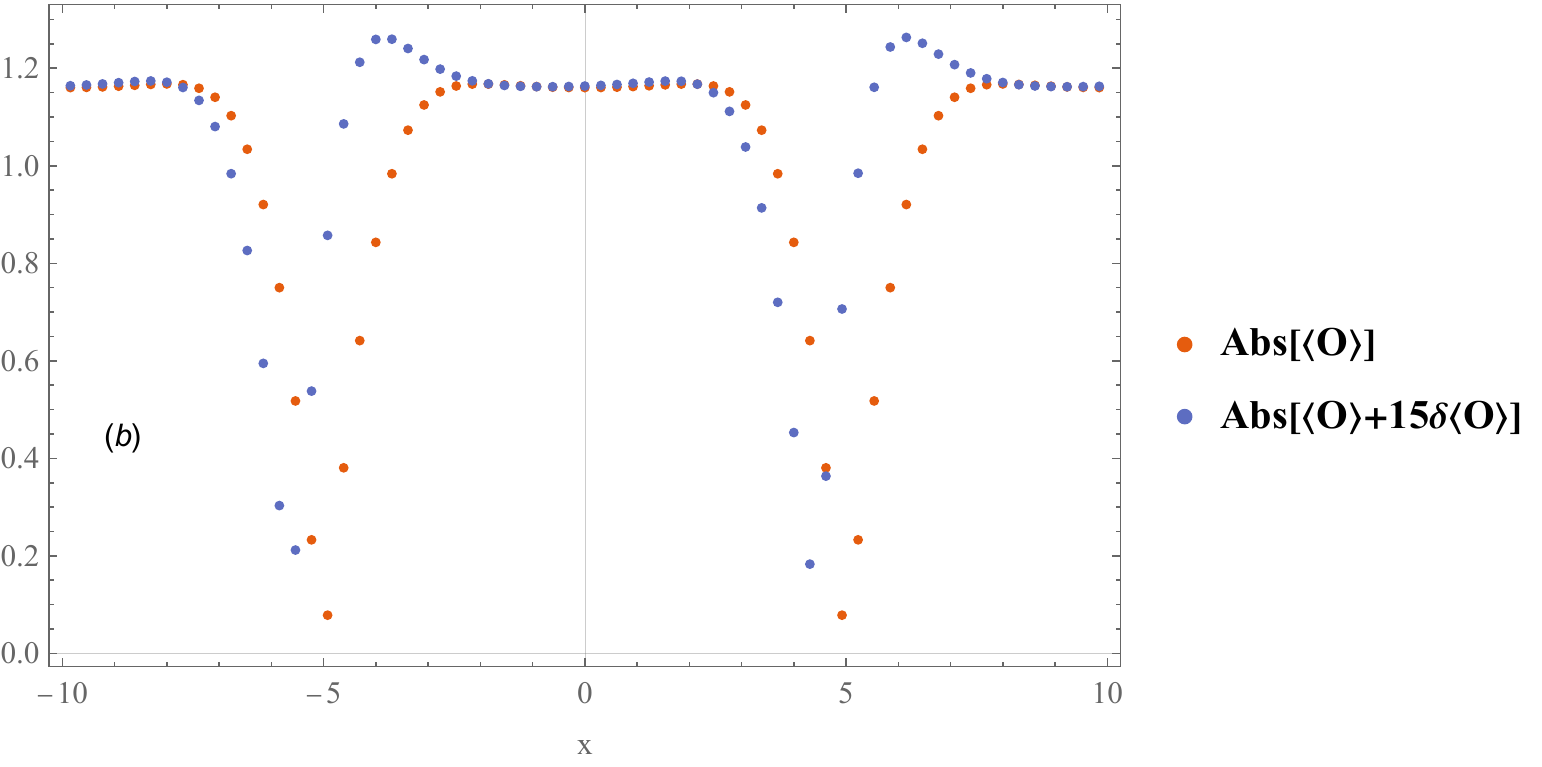}
\includegraphics[scale=0.32]{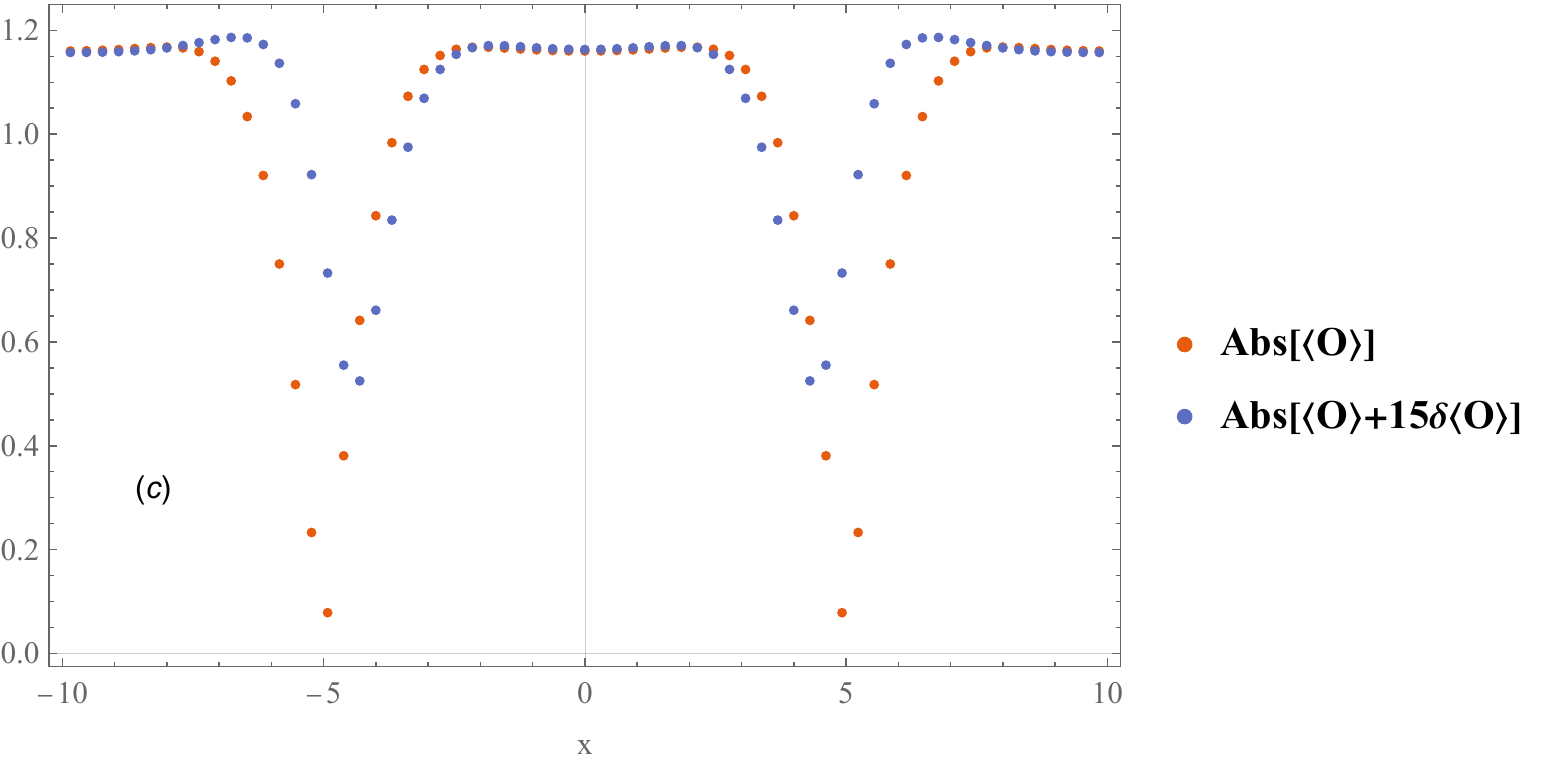}

\caption{One phonon mode and two elastic modes make different influence on
width scale of soliton train at absolute value aspect ($\mu=2.3,L=20,k=0$).
(a) is for phonon mode , (b) is for small elastic mode, (c) is for
large elastic mode.\label{fig:abs value}}

\end{figure}

Fig.\ref{fig:abs value} clarifies the significant influence of three distinct modes on the soliton train width. In Fig. \ref{fig:abs value}(a), the phonon
mode is depicted, revealing that the core position of the soliton train remains unchanged, resulting in an unchanged soliton train width, albeit with amplitude amplification. Moving to Fig. \ref{fig:abs value}(b), the influence of a minor elastic mode is showcased, indicating a deformation in the configuration of the soliton train, leading to a collective leftward shift without altering the core spacing. Remarkably, Fig. \ref{fig:abs value}(c) highlights a crucial observation where the soliton train width contracts due to the proximity of soliton cores induced by the large elastic mode applied to the background solutions. In this sense,  it becomes evident that the impact of the large elastic mode is to bring adjacent solitons close to each other.

Combining the three different features (Fig.\ref{fig:The-phase-argument},
Fig.\ref{fig:imaginary part effect}, Fig.\ref{fig:Real and imaginary} and Fig.\ref{fig:abs value})
of two elastic modes, we can identify their distinctive effects. The large elastic mode primarily governs elastic deformations of the order parameter along the direction of soliton train. However, the small elastic mode is chiefly responsible for phase deformation and presents a grayness oscillation.

\begin{figure}
\includegraphics[scale=0.32]{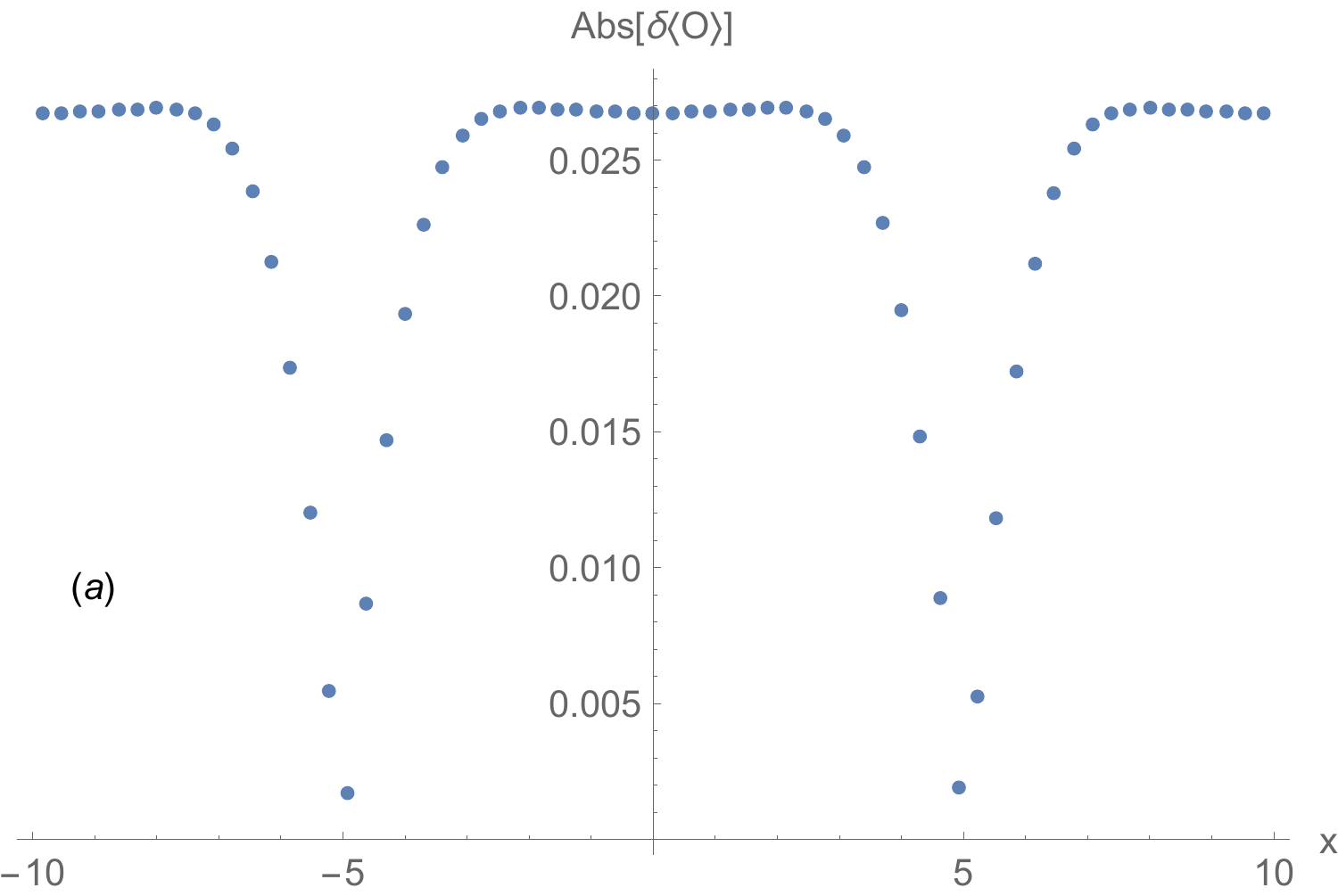}
\includegraphics[scale=0.32]{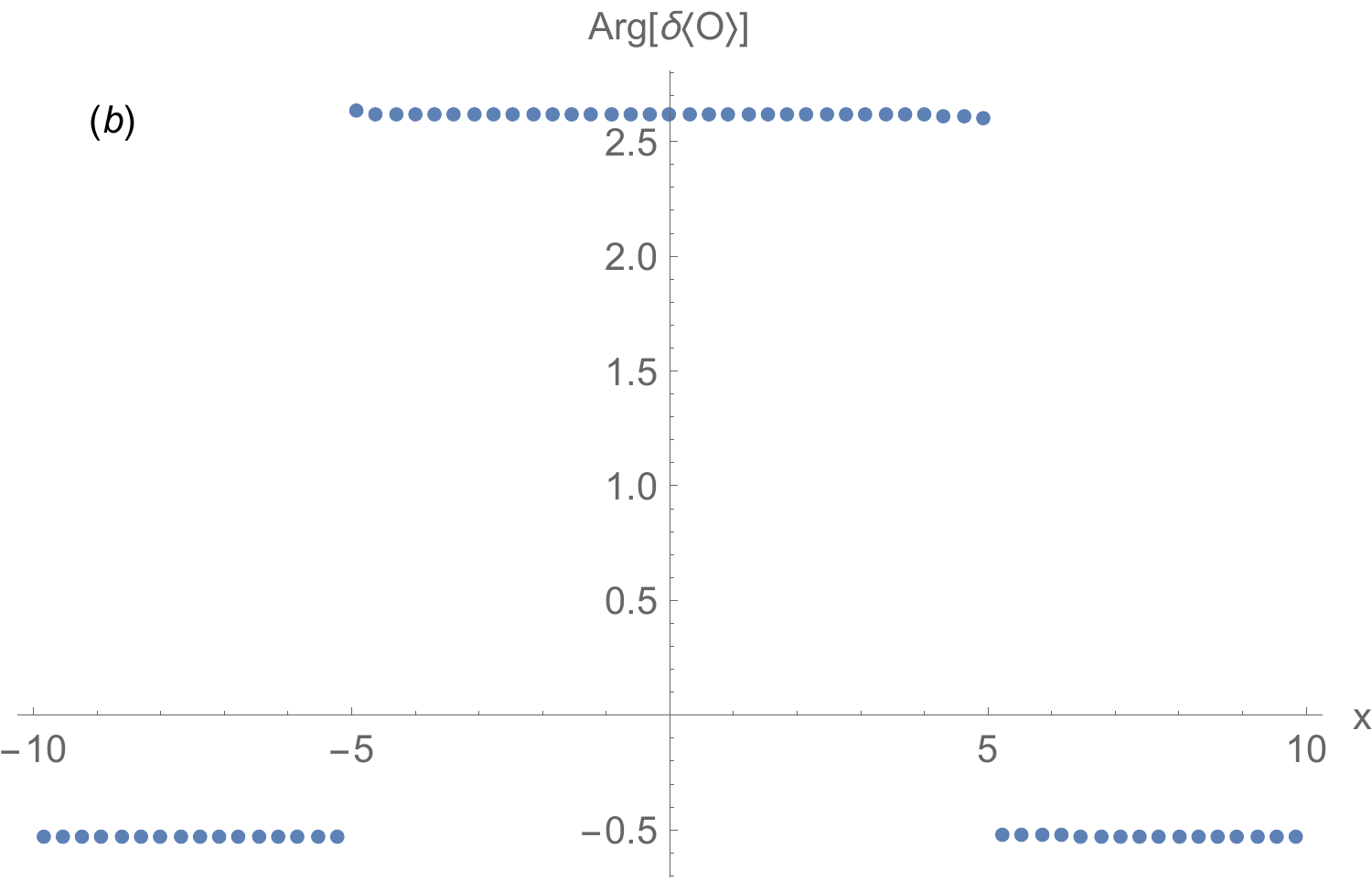}
\includegraphics[scale=0.32]{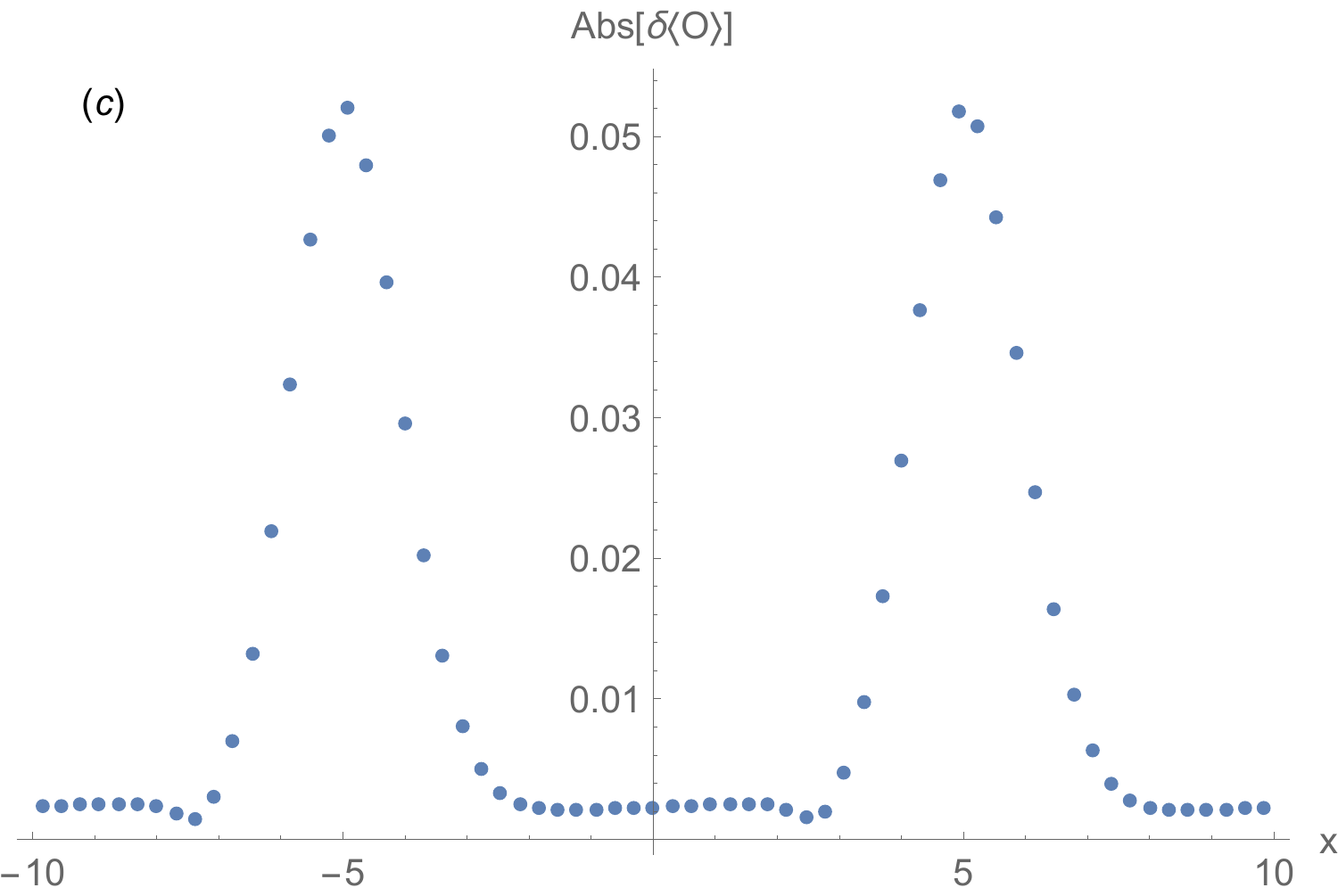}
\includegraphics[scale=0.32]{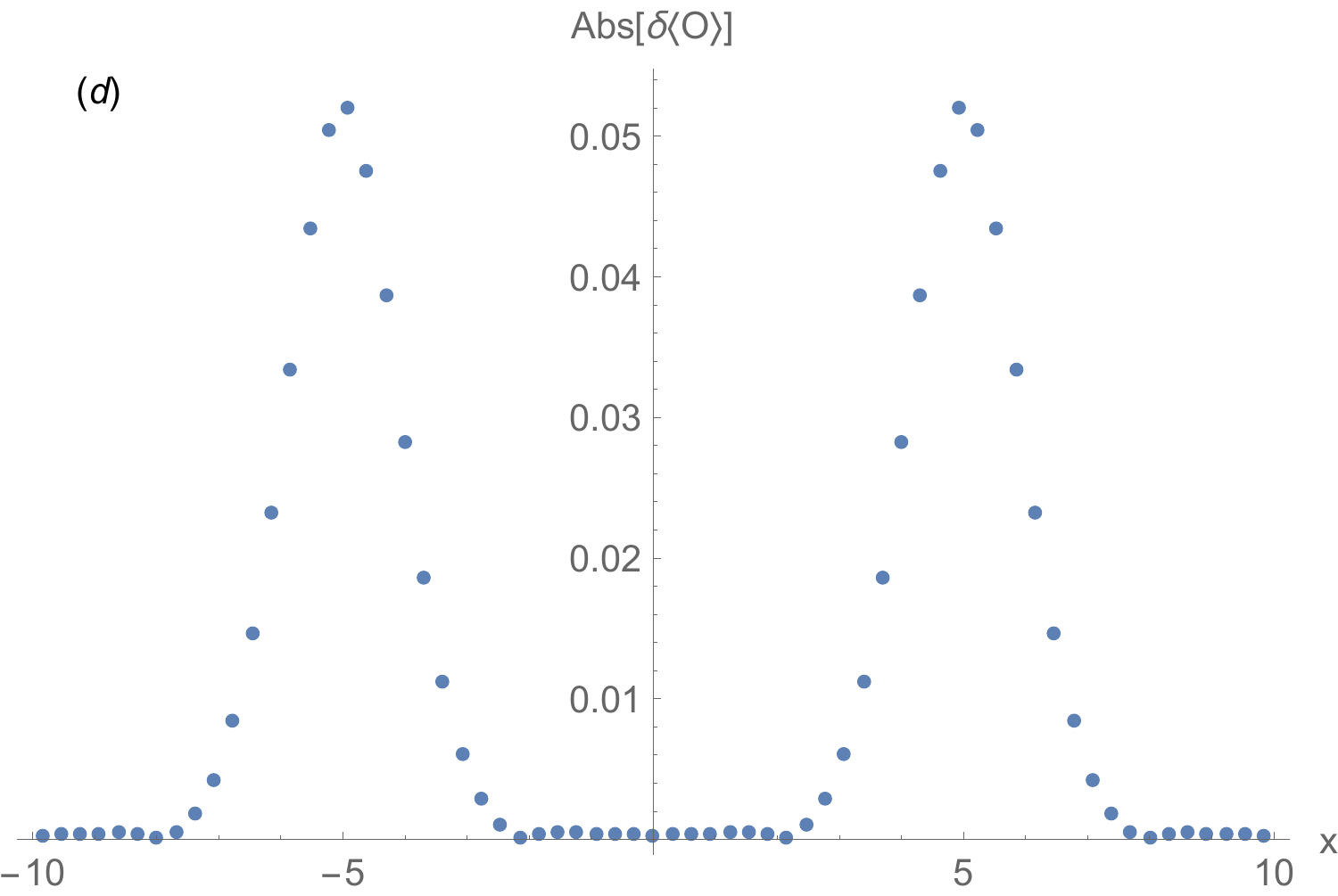}

\caption{The comparison between phonon mode (a) and elastic mode (c and d).
(b): phonon mode phase argument. ($\mu=2.3,L=20,k=0$). From (b),
we can know that the slope of phase argument is zero at uniform part
($-5<x<5$) and this result shows the existence of phonon mode, because
we take $k=0$ to solve QNM here. \label{fig:eigvector}}

\end{figure}

Furthermore, it's worth noting that this phonon mode is gapless. Here, we go back and distinguish between the phonon mode and the elastic mode. As we know, the phase argument for the plane wave
is in direct proportion to the direction of propagation $x$, that is, $Arg\propto k\cdot x$,
the slope $k$ represents the wave number. We depict phonon mode, phase argument and elastic mode in Fig.\ref{fig:eigvector}. Fig. \ref{fig:eigvector}(a)
and Fig. \ref{fig:eigvector}(b) describe the phonon mode and the phase argument of the phonon mode, respectively. It is evident from the plot that the slope remains zero during the uniform section, which coincides with our setting of parameters $k=0$.
Fig. \ref{fig:eigvector}(c) and (d) illustrate small elastic mode and
large elastic mode, respectively. This is discernible because the finite amplitude is primarily concentrated at the soliton core position, with almost zero presence elsewhere.

\begin{figure}
\includegraphics[scale=0.32]{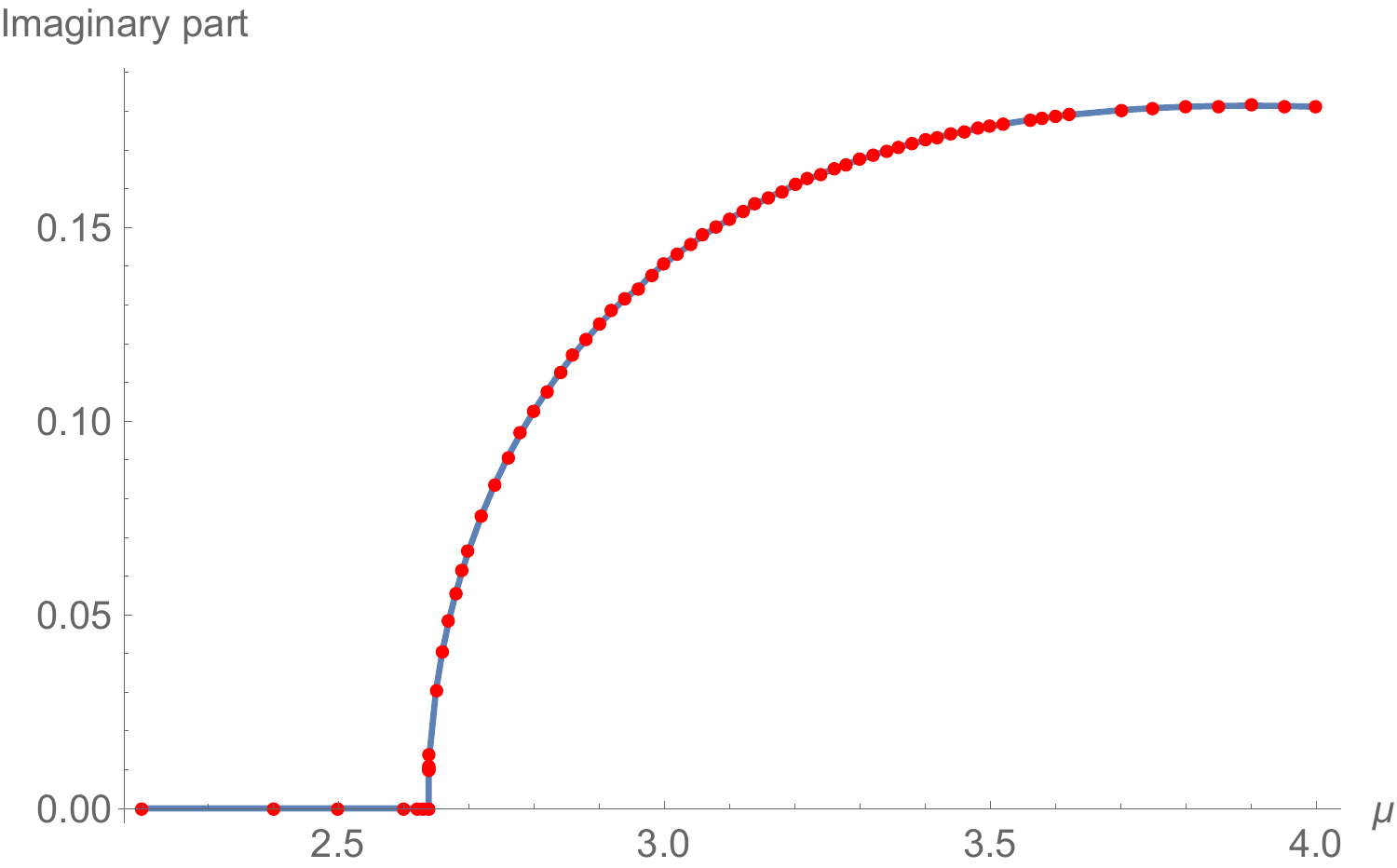}
\includegraphics[scale=0.32]{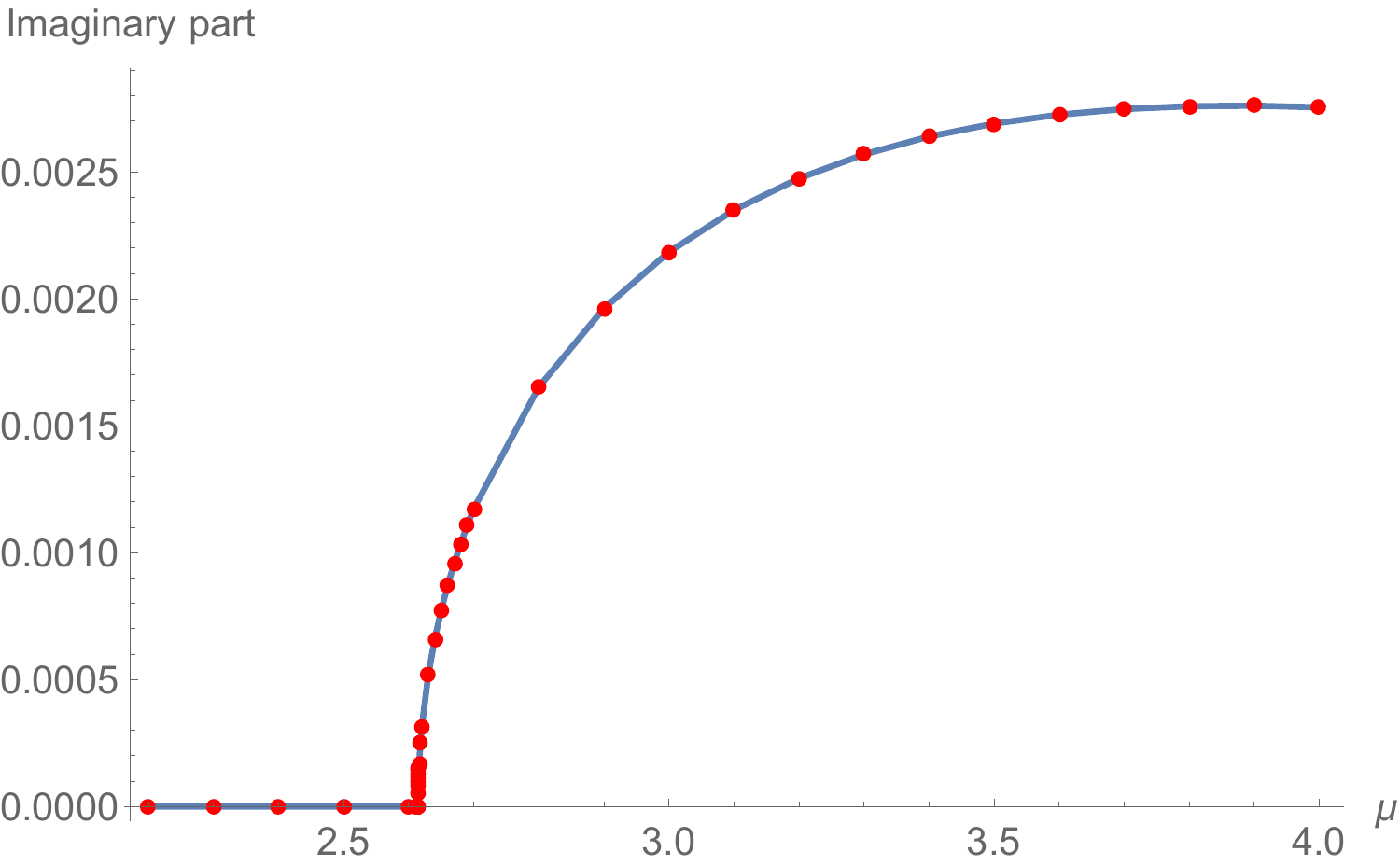}

\caption{The dymamics phase transition for large elastic mode ($k=0$,upper)
and small elastic mode ($k=0.01$,lower) dominated by chemical potential
from stable mode to unstable mode at $L=6$ under standard quantization.
\label{fig:The-dymamics-phase} }
\end{figure}

Remarkably, we observed dynamic phase transition behavior concerning the chemical
potential $\mu$ under the standard quantization framework (BCS-like superfluid system). Setting $L=6$ and monotonously increasing the chemical potential of the system, we identify a critical chemical potential,$\mu_{cl}\simeq2.63882$, beyond which an unstable mode
emerges for $k=0$ situation. This unstable mode manifests as a purely imaginary number and precisely corresponds to the large elastic mode discussed earlier. As the chemical potential $\mu$ increases, the magnitude of this unstable mode increases. We illustrate this phenomenon in Fig.\ref{fig:The-dymamics-phase}.
The presence of this large unstable mode indicates the proximity of adjacent solitons, which ultimately annihilate each other to form a uniform superfluid
\cite{key-16}. In contrast, there is no phase
transition behavior in the alternative quantization scheme, i.e. the soliton
train remains stable within a BEC-like superfluid at zero temperature. Therefore, in this manuscript, we focus only on the dynamic stability of the soliton train under standard quantization.

\begin{figure}
\includegraphics[scale=0.31]{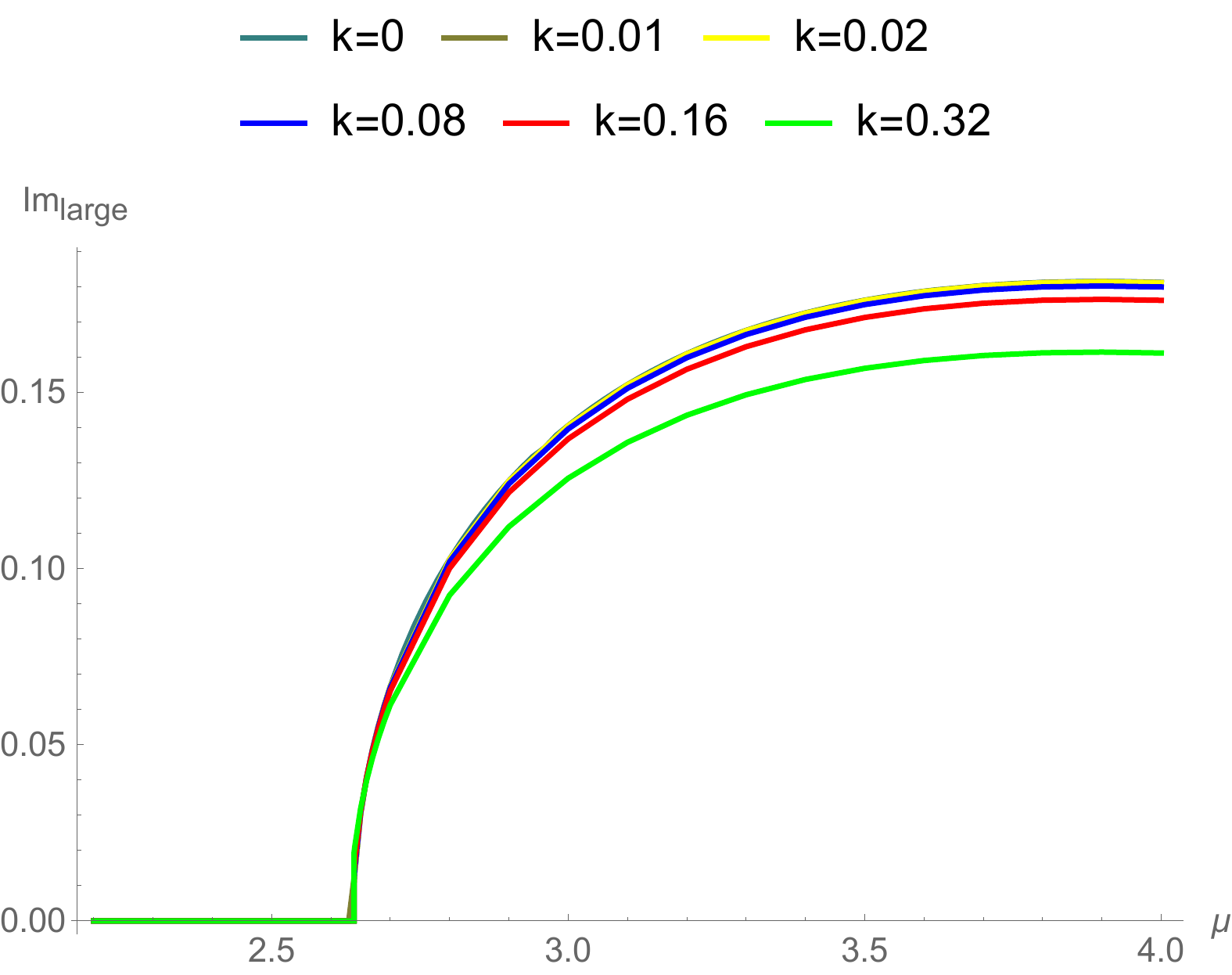}
\includegraphics[scale=0.31]{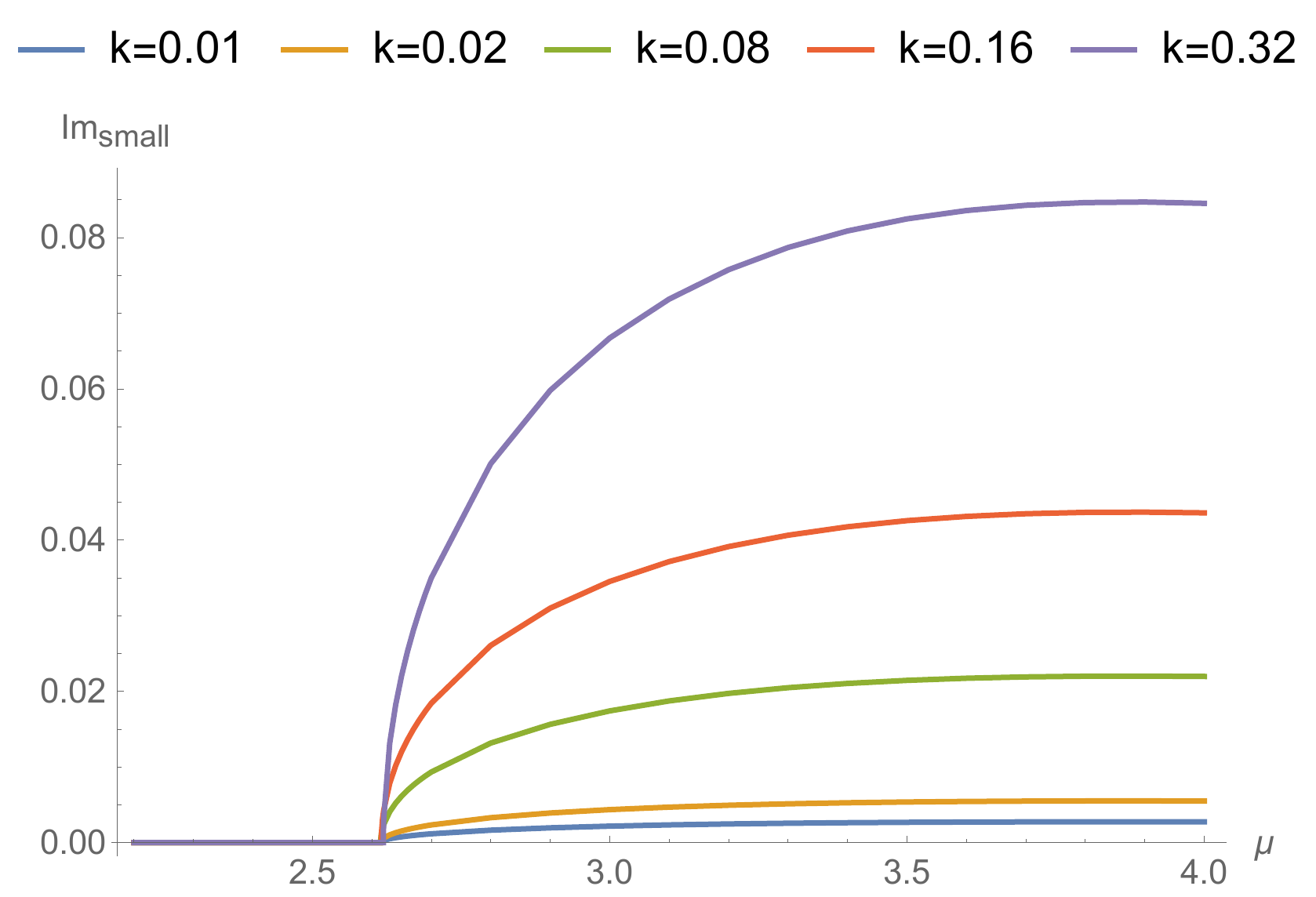}

\caption{The dynamics phase transition for large elastic mode (upper) and small
elastic mode (lower) dominated by chemical potential from stable mode
to unstable mode for different $k$ at $L=6$ under standard quantization
scheme. \label{fig:kmu-im}}

\end{figure}

Moreover, we also observed the other unstable mode, which also presents dynamic phase
transition behaviour as the chemical potential varies (with a
critical chemical potential of $\mu_{cs}=2.6146$)
as $k\neq0$. This unstable mode is exactly the small elastic mode. In summary,
only the large elastic mode becomes unstable beyond critical value
$\mu_{cl}$ for $k=0$, while both the two elastic modes become unstable
for $k\neq0$ as they exceed their respective critical chemical potentials,
which are presented in Fig.\ref{fig:The-dymamics-phase}. The presence of the unstable small elastic mode also makes soliton train become to uniform superfluids. Furthermore, we explore how different perturbation modes $k$ influence the pattern of dynamics phase transitions induced by chemical potential keeping the length scale
$L=6$ fixed, as depicted in Fig.\ref{fig:kmu-im}. It is evident that these modes exhibit a consistent pattern across various wave numbers $k$; however, the magnitude of unstable mode decreases as perturbation
modes $k$ increase for the large elastic mode, whereas it demonstrates an inverse trend for the small elastic mode. Here, $\mathrm{Im_{large}}$ denotes the imaginary part of the large elastic mode, and $\mathrm{Im_{small}}$ signifies the imaginary part of the small elastic mode.

\begin{figure}
\includegraphics[scale=0.28]{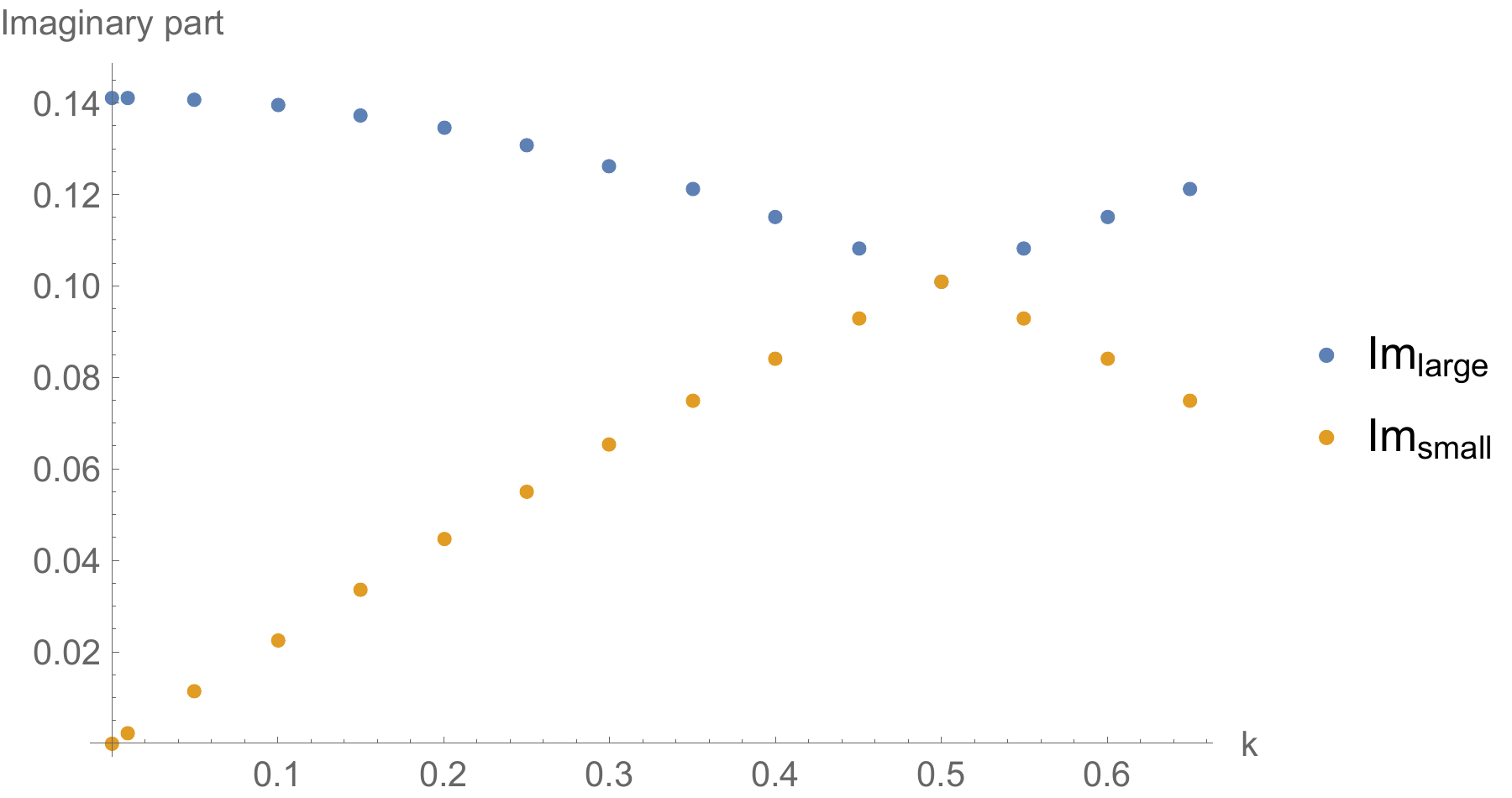}

\caption{The large elastic mode and small elastic mode for different $k$ values. There exists mode degenerate at the first Brillouin zone boundary with wave number $k$ increasing at standard quantization, where $\mu=3.0,L=2\pi$. \label{fig:Mode-crossing}}

\end{figure}

In this study, we delve into the behavior of two types of unstable modes with increasing wave
number $k$, which is shown in Fig.\ref{fig:Mode-crossing}. We observe that small elastic mode exhibits characteristics akin to a hydrodynamic mode. Notably, the two modes degenerate at the first Brillouin zone boundary ($k=0.5$).
Additionally, we indeed observe numerous pairwise degenerate modes
at Brillouin zone boundary.

\begin{figure}
\includegraphics[scale=0.25]{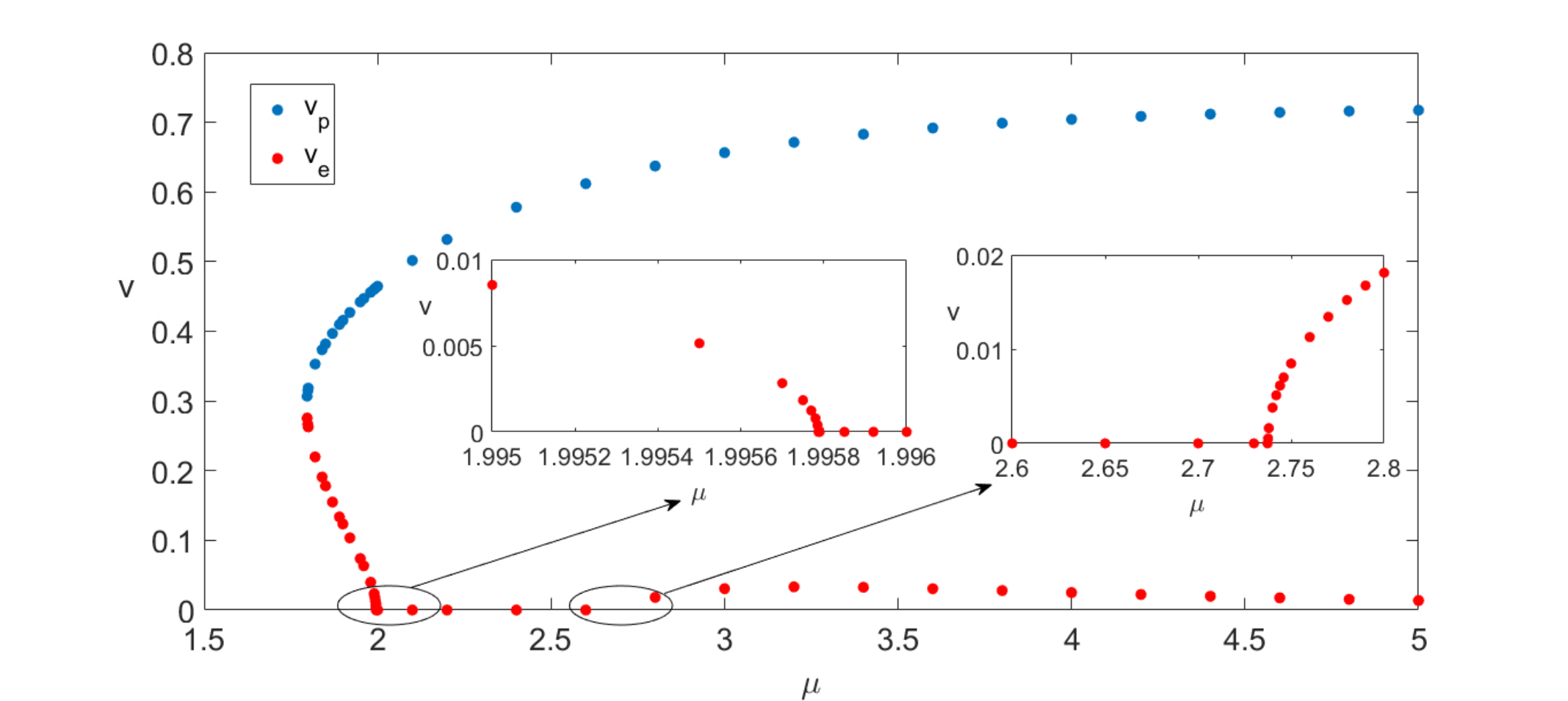}
\includegraphics[scale=0.29]{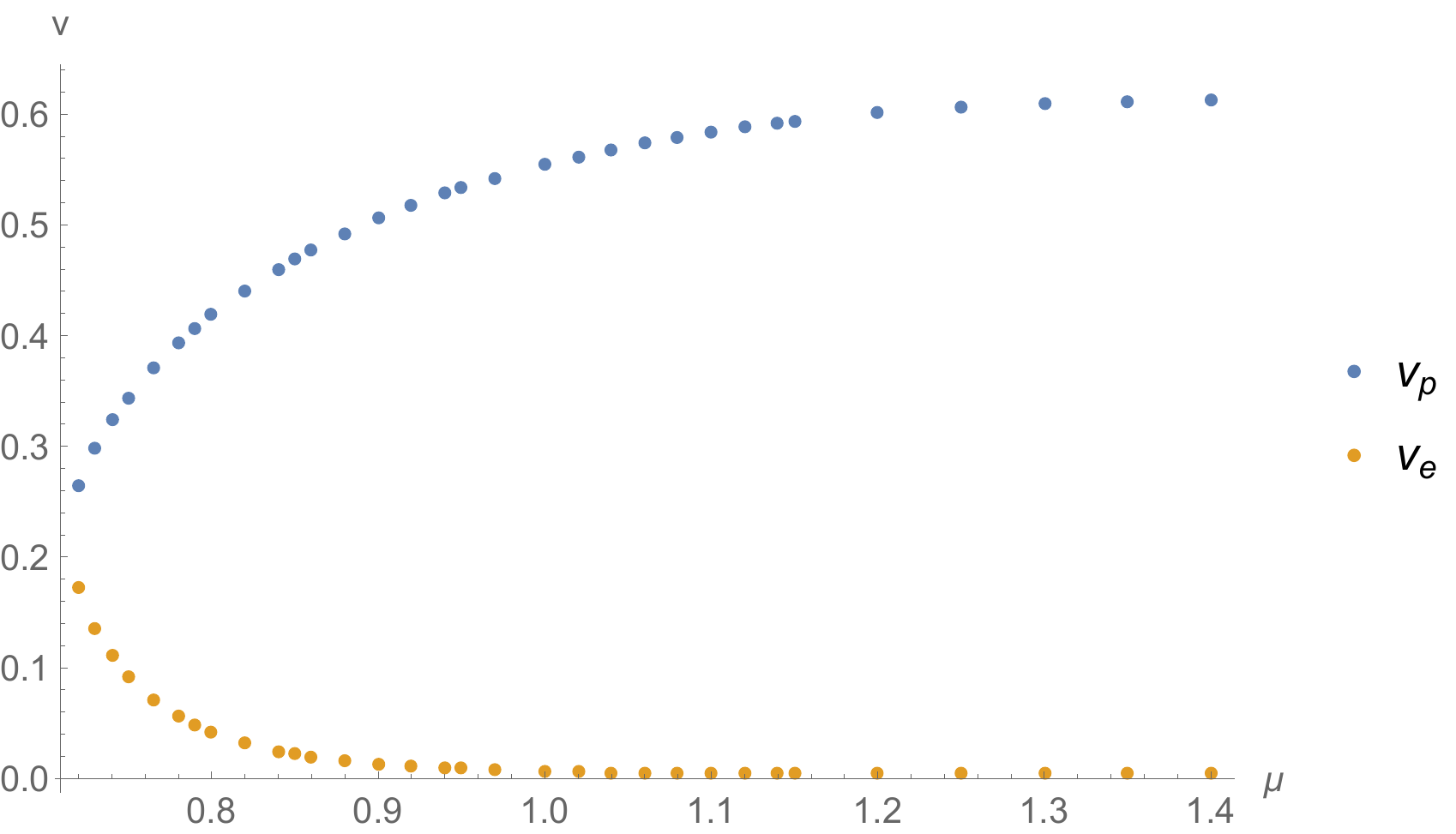}

\caption{The relationship between chemical potential and the two kinds of velocity
(phonon mode as well as gapless elastic mode) on the soliton train
background for standard quantization ($L=12.0$, upper) and alternative
quantization ($L=40$, lower).\label{fig:velocity-mu}}
\end{figure}

Previous research has delved into the relationship between phonon velocity and
chemical potential in homogeneous superfluid context\cite{Guo}. Our current investigation focuses on elucidating this relationship under the framework of the soliton train configuration. The findings are illustrated in Fig.\ref{fig:velocity-mu}. It shows that the velocity of the phonon mode
and the small elastic mode both converge to a constant as the chemical potential attains sufficient magnitude in the two kinds of quantization scheme. Remarkably, as depicted in Fig.\ref{fig:velocity-mu}, there exists a critical range of chemical potential where the velocity of the small elastic mode vanishes, resembling a window of the disappearance of velocity.
Within this range, the small elastic mode is identified as purely imaginary and inherently unstable. We will further investigate this problem in detail and provide a better interpretation in our next work.

Furthermore,  we can identify that the velocity of the phonon mode ($v_{p}$) in
both schemes reaches a constant when the chemical potential becomes sufficiently large, aligning with acoustic speed constraints presented in conformal
fluid theories, where the velocities are $v=\frac{1}{\sqrt{2}},\frac{1}{\sqrt{3}}$ for standard
quantization and alternative quantization, respectively. Accordingly, the speed of the elastic mode also converges towards a constant value as the chemical potential increases.

\section{Summary and discussion}

In this paper, we delve into collective excited modes of a soliton train using the holographic gravity method at zero temperature. We first establish static solutions for the soliton train and then primarily focus on performing linear perturbations
(QNMs) to explore its dynamic behaviour. Through this analysis, we uncover several intriguing results.
At first, we identify three distinct modes: the large elastic mode, the small elastic mode and the phonon mode. Each of these modes exerts a unique influence on the soliton train background. The large mode induces the annihilation of adjacent solitons, leading to the evolution toward a uniform superfluid state, whereas the small mode alters the phase difference between adjacent soliton cores from zero to $\pi$. Additionally, we observe a grayness oscillation phenomenon within the soliton train when the small elastic mode is applied to the system. Furthermore, we investigate the dynamic phase transition regarding the chemical potential parameter. For fixed lattice spacing, the chemical potential plays a pivotal role in determining the stability of the soliton train configuration. Only the large
elastic mode undergoes a phase transition at $k=0$ when it surpasses a critical chemical
potential. However, the two kinds of elastic modes both
become unstable as $k\neq0$. The behavior of the small elastic mode resembles that of a hydrodynamic mode. Moreover, we uncover numerous pairwise-degenerate modes at the Brillouin zone boundary.
Last but not least, we present the relationship between chemical potential and two types of velocity associated with standard quantization and alternative quantization; we observed a chemical potential window for the small elastic mode under standard quantization, where the elastic mode becomes unstable when $L=12$. Consequently, we construct a phase diagram that delineates the dependence of the chemical potential.
In this paper, we study only the impact of the chemical potential factor on the configuration of the soliton train. In the next project, we will study how lattice spacing affects the dynamic behavior of the soliton train system at zero temperature.
Additionally, we anticipate that exploring the system at finite temperature will unveil a plethora of intriguing physical phenomena, enriching our understanding of the multiparticle system.
\begin{acknowledgments}
	This work is partly supported by the National Key Research and Development Program of China with Grant
	No. 2021YFC2203001 as well as the National Natural
	Science Foundation of China with Grant Nos. 12075026, 12035016, 12361141825 and 12375058. Xin Li acknowledges the support form China Scholarship Council (CSC, No. 202008610238). P.Y. acknowledges the support by the China Postdoctoral Science Foundation under Grant Number 2024T170545.
\end{acknowledgments}

\bibliographystyle{JHEP}

\bibliography{references}

\providecommand{\href}[2]{#2}\begingroup\raggedright\begin{thebibliography}{10}

\bibitem{lixin-1}
G.A.~Williams and R.E.~Packard, \emph{{Photographs of quantized vortex lines in rotating He II}}, {\emph{Physical Review Letters} {\bfseries 33} (1974) 280283}.

\bibitem{lixin-2}
E.J.~Yarmchuk, M.J.~Gordon and R.E.~Packard, \emph{{Observation of Stationary Vortex Arrays in Rotating Superfluid Helium}}, {\emph{Physical Review Letters} {\bfseries 43} (1979) 214217}.

\bibitem{collective-13}
Z.~Dutton, M.~Budde, C.~Slowe and L.V.~Hau, \emph{{Observation of Quantum Shock Waves Created with Ultra- Compressed Slow Light Pulses in a Bose-Einstein Condensate}}, {\emph{Science} {\bfseries 293} (2001) 663668}.

\bibitem{key-4}
S.~Donadello, S.~Serafini, M.~Tylutki, L.P.~Pitaevskii, F.~Dalfovo, G.~Lamporesi et~al., \emph{{Observation of Solitonic Vortices in Bose-Einstein Condensates}}, {\emph{Physical Review Letters} {\bfseries 113} (2014) }.

\bibitem{key-5}
Z.~Dutton, M.~Budde, C.~Slowe and L.V.~Hau, \emph{{Observation of Quantum Shock Waves Created with Ultra- Compressed Slow Light Pulses in a Bose-Einstein Condensate}}, {\emph{Science} {\bfseries 293} (2001) 663668}.

\bibitem{key-6}
T.~Yefsah, A.T.~Sommer, M.J.H.~Ku, L.W.~Cheuk, W.~Ji, W.S.~Bakr et~al., \emph{{Heavy solitons in a fermionic superfluid}}, {\emph{Nature} {\bfseries 499} (2013) 426430}.

\bibitem{key-7}
M.J.H.~Ku, B.~Mukherjee, T.~Yefsah and M.W.~Zwierlein, \emph{{Cascade of Solitonic Excitations in a Superfluid Fermi gas: From Planar Solitons to Vortex Rings and Lines}}, {\emph{Phys. Rev. Lett.} {\bfseries 116} (2016) 045304}.

\bibitem{Susskind}
L.~Susskind, \emph{{The world as a hologram}}, {\emph{Journal of Mathematical Physics} {\bfseries 36} (1995) 63776396}.

\bibitem{Maldacena}
J.~Maldacena, \emph{{The Large N Limit of Superconformal Field Theories and Supergravity}}, {\emph{International Journal of Theoretical Physics} {\bfseries 38} (1999) 11131133}.

\bibitem{Witten}
S.~Gubser, I.~Klebanov and A.~Polyakov, \emph{{Gauge theory correlators from non-critical string theory}}, {\emph{Physics Letters B} {\bfseries 428} (1998) 105114}.

\bibitem{key-9}
X.~Li, Y.~Tian and H.~Zhang, \emph{{Generation of vortices and stabilization of vortex lattices in holographic superfluids}}, {\emph{Journal of High Energy Physics} {\bfseries 2020} (2020) }.

\bibitem{key-10}
P.~Yang, X.~Li and Y.~Tian, \emph{{Instability of holographic superfluids in optical lattice}}, {\emph{Journal of High Energy Physics} {\bfseries 2021} (2021) }.

\bibitem{key-11}
Y.-K.~Yan, S.~Lan, Y.~Tian, P.~Yang, S.~Yao and H.~Zhang, \emph{{Holographic dissipation prefers the Landau over the Keldysh form}},  2023.

\bibitem{key-12}
P.~Wittmer, C.-M.~Schmied, T.~Gasenzer and C.~Ewerz, \emph{{Vortex Motion Quantifies Strong Dissipation in a Holographic Superfluid}}, {\emph{Physical Review Letters} {\bfseries 127} (2021) }.

\bibitem{Guo}
M.~Guo, S.~Lan, C.~Niu, Y.~Tian and H.~Zhang, \emph{{Note on zero temperature holographic superfluids}}, {\emph{Classical and Quantum Gravity} {\bfseries 33} (2016) 127001}.

\bibitem{gaomeng}
M.~Gao, Y.~Jiao, X.~Li, Y.~Tian and H.~Zhang, \emph{{Black and gray solitons in holographic superfluids at zero temperature}}, {\emph{Journal of High Energy Physics} {\bfseries 2019} (2019) }.

\bibitem{key-15}
D.A.~Takahashi, \emph{{Bogoliubov--de Gennes soliton dynamics in unconventional Fermi superfluids}}, {\emph{Phys. Rev. B} {\bfseries 93} (2016) 024512}.

\bibitem{Joshua}
J.E.~Rothenberg and H.K.~Heinrich, \emph{{Observation of the formation of dark-soliton trains in optical fibers}}, {\emph{Opt. Lett.} {\bfseries 17} (1992) 261263}.

\bibitem{KEStrecker}
K.E.~Strecker, G.B.~Partridge, A.G.~Truscott and R.G.~Hulet, \emph{{Formation and propagation of matter-wave soliton trains}}, {\emph{Nature} {\bfseries 417} (2002) 150153}.

\bibitem{Nguyen}
J.H.V.~Nguyen, D.~Luo and R.G.~Hulet, \emph{{Formation of matter-wave soliton trains by modulational instability}}, {\emph{Science} {\bfseries 356} (2017) 422426}.

\bibitem{Nishioka}
T.~Nishioka, S.~Ryu and T.~Takayanagi, \emph{{Holographic superconductor/insulator transition at zero temperature}},  2010.

\bibitem{M_Randeria}
M.~Randeria, \emph{{In Bose-Einstein Condensation}}, Cambridge University Press (1995).

\bibitem{Timmermans}
E.~Timmermans, K.~Furuya, P.W.~Milonni and A.K.~Kerman, \emph{{Prospect of creating a composite Fermi–Bose superfluid}}, {\emph{Physics Letters A} {\bfseries 285} (2010) 228233}.

\bibitem{key-16}
S.~Dutta and E.J.~Mueller, \emph{{Collective Modes of a Soliton Train in a Fermi Superfluid}}, {\emph{Phys. Rev. Lett.} {\bfseries 118} (2017) 260402}.

\bibitem{S.A}
S.A.~Hartnoll, C.P.~Herzog and G.T.~Horowitz, \emph{{Building a Holographic Superconductor}}, {\emph{Phys. Rev. Lett.} {\bfseries 101} (2008) 031601}.

\bibitem{C.P}
S.A.~Hartnoll, C.P.~Herzog and G.T.~Horowitz, \emph{{Holographic superconductors}}, {\emph{Journal of High Energy Physics} {\bfseries 2008} (2008) 015015}.

\bibitem{Guo2}
M.~Guo, C.~Niu, Y.~Tian and H.~Zhang, \emph{{Modave Lectures on Applied AdS/CFT with Numerics}},  2016.

\bibitem{Kovtun}
C.P.~Herzog, P.K.~Kovtun and D.T.~Son, \emph{{Holographic model of superfluidity}}, {\emph{Physical Review D} {\bfseries 79} (2009) }.

\bibitem{Klebanov}
I.R.~Klebanov and E.~Witten, \emph{{Ads/CFT correspondence and symmetry breaking}}, {\emph{Nuclear Physics B} {\bfseries 556} (1999) 89–114}.

\bibitem{Freedman}
P.~Breitenlohner and D.Z.~Freedman, \emph{{Positive Energy in anti-De Sitter Backgrounds and Gauged Extended Supergravity}}, {\emph{Physics Letters B} {\bfseries 115} (1982) 197201}.

\end{thebibliography}\endgroup

\end{document}